\documentclass[lettersize,journal]{IEEEtran}
\usepackage{amsmath,amsfonts}
\usepackage{array}
\usepackage[caption=false,font=normalsize,labelfont=sf,textfont=sf]{subfig}
\usepackage{textcomp}
\usepackage{stfloats}
\usepackage{url}
\usepackage{verbatim}
\usepackage{graphicx}
\def\BibTeX{{\rm B\kern-.05em{\sc i\kern-.025em b}\kern-.08em
		T\kern-.1667em\lower.7ex\hbox{E}\kern-.125emX}}
\usepackage{balance}

\ifCLASSOPTIONcompsoc
\usepackage[nocompress]{cite}
\else

\usepackage{cite}
\fi

\ifCLASSINFOpdf

\else

\fi
\usepackage{cite}
\usepackage{amsmath,amssymb,amsfonts}
\usepackage{graphicx}
\usepackage{textcomp}
\usepackage{xcolor}
\usepackage{cite} 
\usepackage{url}
\usepackage{epstopdf}
\usepackage{subcaption}
\usepackage[colorlinks,
linkcolor=black,       
anchorcolor=black,  
citecolor=black,
]{hyperref}
\usepackage{mathrsfs}
\usepackage{booktabs} 
\usepackage{tabularx} 
\usepackage{array}    
\usepackage{rotating}
\usepackage{tabulary}
\usepackage{lipsum} 
\usepackage{tabularx} 

\usepackage{algorithm}
\usepackage{algpseudocode}  
\usepackage{amsmath}

\usepackage{listings}
\usepackage{xcolor} 

\usepackage{amssymb}
\usepackage{pifont}  
\usepackage{tabularx}
\usepackage{booktabs}

\usepackage{graphicx}

\usepackage{textcomp}
\usepackage{amssymb}
\usepackage[percent]{overpic}

\usepackage{graphicx}
\usepackage{hyperref}
\usepackage[percent]{overpic}

\usepackage{tikz}
\usepackage{pgfplots}
\usepackage{pgfplotstable} 

\newcommand{\cmark}{\ding{51}} 
\newcommand{\xmark}{\ding{55}} 

\usepackage{pgfplots}
\pgfplotsset{compat=1.17}
\usepackage{tikz}

\def\BibTeX{{\rm B\kern-.05em{\sc i\kern-.025em b}\kern-.08em
T\kern-.1667em\lower.7ex\hbox{E}\kern-.125emX}}

\begin{document}

\title{Lessons from Deploying Learning-based CSI Localization on a Large-Scale ISAC Platform}

\author{Tianyu~Zhang,
	Dongheng~Zhang,
	Ruixu~Geng,
	Xuecheng~Xie,
	Shuai~Yang,
	and~Yan~Chen*,~\IEEEmembership{Senior Member,~IEEE}
	\IEEEcompsocitemizethanks{\IEEEcompsocthanksitem T. Zhang, D. Zhang, R. Geng, X. Xie, S. Yang, Y. Chen are with the School of Cyber Science and Technology, University of Science and Technology of China, Hefei 230026, China
		(E-mail: tianyuzhang@mail.ustc.edu.cn, dongheng@ustc.edu.cn, {gengruixu, xuechengxie, ys1025}@mail.ustc.edu.cn, eecyan@ustc.edu.cn).
		\IEEEcompsocthanksitem T. Zhang is an intern at the Institute of Artificial Intelligence, Hefei Comprehensive National Science Center, Hefei, China.
		\IEEEcompsocthanksitem Corresponding author: Yan Chen (E-mail: eecyan@ustc.edu.cn)}
}

\IEEEtitleabstractindextext{%
	\begin{abstract}
		In recent years, Channel State Information (CSI), recognized for its fine-grained spatial characteristics, has attracted increasing attention in WiFi-based indoor localization. 
		However, despite its potential, CSI-based approaches have yet to achieve the same level of deployment scale and commercialization as those based on Received Signal Strength Indicator (RSSI).  
		A key limitation lies in the fact that most existing CSI-based systems are developed and evaluated in controlled, small-scale environments, limiting their generalizability. 
		To bridge this gap, we explore the deployment of a large-scale CSI-based localization system involving over 400 Access Points (APs) in a real-world building under the Integrated Sensing and Communication (ISAC) paradigm. 
		We highlight two critical yet often overlooked factors: the underutilization of unlabeled data and the inherent heterogeneity of CSI measurements. 
		To address these challenges, we propose a novel CSI-based learning framework for WiFi localization, tailored for large-scale ISAC deployments on the server side. 
		Specifically, we employ a novel graph-based structure to model heterogeneous CSI data and reduce redundancy. 
		We further design a pretext pretraining task that incorporates spatial and temporal priors to effectively leverage large-scale unlabeled CSI data. 
		Complementarily, we introduce a confidence-aware fine-tuning strategy to enhance the robustness of localization results. 
		In a leave-one-smartphone-out experiment spanning five floors and \(25,600 \, m^2\), we achieve a median localization error of 2.17 meters and a floor accuracy of 99.49\%. 
		This performance corresponds to an 18.7\% reduction in mean absolute error (MAE) compared to the best-performing baseline.
	\end{abstract}
}

\maketitle
\IEEEdisplaynontitleabstractindextext

\IEEEpeerreviewmaketitle

\section{Introduction}

\IEEEPARstart{I}ndoor localization techniques have been the subject of extensive research over the past two decades~\cite{hu2022experience}.
WiFi, in particular, has emerged as the most promising technology for this purpose, largely due to its widespread infrastructure~\cite{ni2022experience}. 
If the existing WiFi infrastructure can be utilized for location-based services, there is no need to deploy additional hardware.  
Early research on WiFi localization primarily focused on methods that use the Received Signal Strength Indicator (RSSI), a metric that characterizes the attenuation of radio signals during propagation~\cite{yang2013rssi}. 
In recent years, several large-scale, data-driven indoor localization systems relying on RSSI have been successfully deployed, providing daily location services to millions of users across multiple cities~\cite{ni2022experience, hu2022experience, guo2022wepos}. 

Recent Channel State Information (CSI), which captures detailed attenuation and phase shift information at the granularity of Orthogonal Frequency Division Multiplexing (OFDM) subcarriers~\cite{yang2013rssi}, has emerged as a promising alternative due to superior localization accuracy and robustness~\cite{liu2019survey, vasisht2016decimeter, ayyalasomayajula2020deep, zhang2022toward, zhang2024rloc}. 
However, compared to the commercialization progress and deployment scale of RSSI methods, CSI-based methods have not yet reached the same level of maturity. 
A major limitation of existing CSI-based data-driven systems is generally developed and evaluated in specialized deployments with controlled conditions, making them less applicable to real-world applications, as indicated in Tab.~\ref{tab:comparison}. 
These systems are designed and evaluated in small-scale deployments, while assuming that the CSI structure is uniform. 
To bridge this gap, we explore a CSI-based, learning-driven localization system within a large-scale ISAC (Integrated Sensing and Communication) platform, comprising over 400 Access Points (APs) operating in a building. 
This platform operates under the centralized control of WLAN controllers to support daily communication needs while simultaneously enabling CSI acquisition for sensing and localization. 
Notably, it adopts a server-side architecture\footnote{Ethical Concern: We prioritize user privacy. No personally identifiable information (PII) was collected or used in this study. We do not correlate a user’s location with their true identity.} in which data collection is performed without requiring any modifications to or active participation from client devices~\cite{jaisinghani2018experiences}.
Building upon this practically deployed platform, we further identify two critical aspects that are often overlooked by existing CSI-based, data-driven systems. 


\begin{table*}[h]
	\centering
	\caption{Comparison of Recent CSI-Based Learning-based Localization Methods and Our Deployment}
	\label{tab:comparison}
	\resizebox{\textwidth}{!}{%
		\begin{tabularx}{\textwidth}{l c c c c c}
			\toprule
			\textbf{System} & \textbf{APs} & \textbf{AP Types} & \textbf{CSI Dim.} & \textbf{Unlab. Data} & \textbf{Hetero. CSI} \\
			\midrule
			ConFi~\cite{chen2017confi} & 3 & Intel 5300 & \(3 \times 1 \times 30\) & \xmark & \xmark \\
			DLM~\cite{arnold2019novel} & 1 & USRP & \(64 \times 1 \times 922\) & \xmark & \xmark \\
			DLoc~\cite{ayyalasomayajula2020deep} & 3 or 4 & Quantenna APs & \(4 \times 1 \times 216\) & \xmark & \xmark \\
			RLoc~\cite{zhang2024rloc} & 3 or 4 & Intel 5300 & \(3 \times 1 \times 30\) & \xmark & \xmark \\
			MSG~\cite{liu2025graph} & 4 & Intel 5300 & \(3 \times 1 \times 30\) & \xmark & \xmark \\
			\midrule
			\textbf{Our Deployment} & \textbf{44 or 436} & \textbf{H3C WA6520, WA6526E, WA6530} & \textbf{Over 6 unique configurations} & \cmark & \cmark \\
			\bottomrule
		\end{tabularx}
	}
\end{table*}


Firstly, server-side ISAC systems in continuous operation naturally generate large volumes of unlabeled data, which present valuable opportunities for exploitation. 
Traditional CSI-based data-driven localization systems~\cite{chen2017confi, arnold2019novel, ayyalasomayajula2020deep, zhang2024rloc, liu2025graph} have typically divided the process into two distinct phases: offline and online. 
In the offline phase, a fingerprint dataset or learning-based model is constructed using data collected from the survey site, mapping WiFi signals to locations. 
In the online phase, new signals are used alongside the fingerprint database or learning-based model to determine the location. 
However, recent research in fields such as computer vision~\cite{chen2020simple}, natural language processing~\cite{devlin2018bert}, and wireless sensing~\cite{song2022rf, li2022unsupervised} highlights the potential of unlabeled data. 
These methods provide evidence that leveraging pretraining strategies with unlabeled data enhances the performance of models trained with limited labeled data~\cite{chen2020simple, oquab2023dinov2}. 
This insight aligns closely with the data availability pattern observed in operational ISAC platforms, where a continuous stream of nearly cost-free unlabeled data coexists with limited and expensive labeled data. 
These factors drive us to explore how unlabeled data can be integrated into the offline phase to improve localization performance. 





Secondly, CSI signals may be heterogeneous in ISAC systems, which complicates the encoding process for the learning-based model input. 
In previous works, the equipment used in experimental settings ensured uniformly received CSI signals.
Traditional systems process or encode WiFi signals in the form of fixed-length vectors or matrices~\cite{chiu2025graph}. 
In contrast, CSI signals in ISAC systems exhibit heterogeneous characteristics, posing challenges for traditional systems. 
First, human movement and power limitations lead to interactions with different APs. 
Not all signals from the APs at a given location may be fully scanned, leaving several entries in the vectors or matrices empty. This approach may suffer from the missing-value problem~\cite{zhuo2022grafics}. 
Second, the ISAC platform typically features a mix of communication configurations~\cite{he2023sencom}. 
For instance, user devices equipped with MIMO (Multiple Input Multiple Output) technologies adjust their modes between diversity and multiplexing based on communication needs~\cite{he2023sencom}. 
This leads to variations in the length of the CSI dimension, which results in dimensional issues for vectors or matrices encoding. 
Moreover, in a large-scale deployment, these APs of different types may have varying numbers of antennas, further contributing to dimensional issues.
These factors motivate us to explore an encoding approach suited for heterogeneous CSI data.

Thus, in this paper, we propose a novel deep learning framework for WiFi localization in large-scale ISAC deployments from a server-side perspective. 
We focus on three main design aspects:
\textbf{(i) Incorporating pretraining techniques into the data-driven framework.}
Recent work~\cite{chen2020simple, oquab2023dinov2, devlin2018bert, song2022rf, li2022unsupervised, song2024unleashing, fang2024prism, salihu2024self, zhang2025umimo} have demonstrated strong performance in their respective tasks, however they are not directly applicable to localization problems, due to fundamental differences in signal characteristics and specific proxy tasks~\cite{li2022unsupervised}. 
To address this gap, we leverage both time~\cite{ferrand2021triplet} and spatial dimensions~\cite{taner2023channel} of prior knowledge to design proxy tasks for pretraining.
We use a contrastive prediction task to group temporally close samples from the same transmitter, which generally correspond to geographically proximate samples.
Then, we incorporate a triplet loss to integrate signal strength and AP locations, providing a coarse relative localization logic for metric learning.
These tasks are performed in a multi-task pretraining framework through branch networks.
\textbf{(ii) Graph encoding techniques for heterogeneous WiFi signals.} 
Several recent studies~\cite{wang2018improved, zheng2021grafin, lezama2021indoor, sun2021novel, lezama2023application, wang2024graph, liu2025graph, chen2023graph, chiu2025graph}  utilize a graph to model the RF signals, offering valuable insights. 
However, these methods primarily represent APs as nodes, utilizing homogeneous RSSI or CSI features as node attributes, or modeling location as a node with fixed-vector features, which overlooks both the heterogeneity of CSI dimensions and the issue of missing values. 
To address this gap, we define the minimal node as the transmitter-receiver antenna pair and extend it to accommodate various transmitter and receiver antenna configurations. 
Moreover, we incorporate AP location features, similar to an AP ID, into the node, which avoids the encoding of missing values. 
\textbf{(iii) Incorporating Confidence-Aware Fine-Tuning into Location Estimation.} 
Though graph encoding is effective for encoding heterogeneous CSI information, it may not directly apply to our scenario.
Heterogeneous information is more likely to violate the independent and identically distributed (i.i.d.) assumptions~\cite{vovk2005algorithmic} compared to homogeneous signals.
For instance, the issue arises when the user device operates in diversity mode during site surveys, while using multiplexing in the test set, which would never occur in a homogeneous setup. 
To mitigate this issue, we introduce a confidence-aware fine-tuning phase~\cite{lakshminarayanan2017simple,he2019bounding} to the pretraining model, where we model both the predicted location and the associated confidence level. 
This confidence estimation reduces the black-box nature of model outputs and, on the other hand, can be used for weighting to enhance the robustness of localization results. 

We built GLow, a \underline{G}raph neural network-based \underline{lo}calization model that leverages \underline{W}iFi signal data, and deployed it in two indoor large-scale scenarios, encompassing a total of 70,000 data points: the first covers approximately \(4000 \, \text{m}^2\) and involves data collected from five types of smartphones; the second spans approximately \(25,600 \, \text{m}^2\) across five floors, utilizing seven types of smartphones. 
We summarize our results below:
\begin{itemize}
	\item The performance of GLow exceeds that of state-of-the-art data-driven localization systems~\cite{torres2015comprehensive, jang2018indoor, arnold2019novel, liu2025graph}.
	\item The performance of pretraining techniques demonstrates its advantages in localization tasks when compared with state-of-the-art baselines~\cite{you2020graph, hou2022graphmae, ni2022experience, ferrand2021triplet, taner2023channel}. 
	\item We conduct extensive ablation studies to analyze the effectiveness of spatiotemporal pretraining and confidence-aware finetuning in the GLow model. 
   \item Our pretraining method facilitates better temporal generalization in localization and sustains performance improvements in fine-tuning tasks even after a half-year. 
	\item We conduct a leave-one-smartphone-out experiment spanning five floors and \(25,600 \, m^2\), where GLow achieves a median error of 2.17m, surpassing the baseline MAE by 18.7\%, and attains a floor accuracy of 99.49\%.
\end{itemize}

\textbf{Our contributions are summarized below}:

\begin{itemize}
	\item We propose GLow, a novel CSI-based learning framework for WiFi localization, specifically designed for large-scale ISAC deployments from a server-side perspective. Our approach incorporates spatiotemporal pretraining and confidence-aware fine-tuning to deliver state-of-the-art localization performance. 
	\item GLow provides an approach that leverages the graph-based structure to encode the input and effectively address the issues of heterogeneous signals. 
	\item GLow leverages server-side prior knowledge to efficiently utilize large-scale unlabeled data, demonstrating the role of pretraining in enhancing the localization task. 
	\item To the best of our knowledge, GLow is the first CSI-based localization framework designed and evaluated at a scale involving more than 400 APs. Extensive experiments demonstrate its potential as a prototype localization system for large-scale server-side ISAC deployments. 
\end{itemize}

The organization of our paper is as follows: In Sec. \ref{section:work}, we summarize recent works in the field.
In Sec. \ref{section:formulation}, we formulate the localization problem and demonstrate the heterogeneous characteristics of WiFi signals using two toy examples. 
Sec. \ref{section:system} introduces our system methodology. In Sec. \ref{section:dataset}, we describe the dataset used. Sec. \ref{section:evaluation} presents the evaluation results of our method. Finally, we analyze the limitations in Sec. \ref{section:limitations} and conclude the paper in Sec. \ref{section:conclusion}.

\section{RELATED WORK}
\label{section:work}
To provide a clear overview, we discuss the related work from four perspectives:

\textbf{Wireless Signal Representation Learning}:
Advancements in representation learning across domains such as computer vision~\cite{chen2020simple, oquab2023dinov2} and natural language processing~\cite{devlin2018bert} have catalyzed innovative applications in wireless signal processing~\cite{song2022rf, li2022unsupervised, song2024unleashing, fang2024prism, zhang2025umimo, salihu2024self}. 
Although these methods have demonstrated strong performance in their respective tasks, they are not directly applicable to localization problems due to fundamental differences in signal characteristics~\cite{li2022unsupervised}. 
In this context, our approach prioritizes localization-aware representation learning by leveraging spatial and temporal priors available on the server side. 
Although~\cite{salihu2024self} focuses on learning general-purpose channel features robust to fading and system impairments by leveraging fading characteristics, our work does not conflict with this research. We place greater emphasis on exploiting location-specific priors. 


\textbf{Leveraging Unlabeled Data in Wireless localization}:
Several studies have explored the use of unlabeled datasets in wireless localization.
A common approach in prior research is fingerprint database updating~\cite{chen2020fido, li2024train, wang2023automatic, zhang2023domain, chiu2025graph}, which leverages new unlabeled datasets to enhance existing fingerprints.
These systems typically prioritize training with labeled data before incorporating unlabeled data. 
Other studies in localization have explored semi-supervised methods~\cite{ferrand2021triplet, ghazvinian2021modality, karmanov2021wicluster, taner2023channel}, which leverage both unlabeled and labeled datasets for training, thereby reducing the cost of manual site surveys. 
Those works explore the semi-supervised framework by utilizing both unlabeled and labeled datasets to train the localization model. 
Our approach is inspired by these works, but unlike the aforementioned approaches, we follow a pretraining-and-finetuning paradigm: we first pretrain the model solely on unlabeled data, and then fine-tune it using labeled data for downstream localization tasks.

\textbf{CSI-based Wireless Localization Systems}: 
CSI-based localization techniques can be broadly classified into three categories: angle-based~\cite{zhang2022toward, yang2023multiple, zhang2024rloc}, range-based~\cite{vasisht2016decimeter, zhang2020peer}, and data-driven~\cite{ayyalasomayajula2020deep}. 
Angle-based methods, which utilize Array Signal Processing (ASP) to estimate the Angle of Arrival (AoA), and range-based methods, which often rely on techniques like frequent channel dropping, both face challenges related to antenna spacing and communication requirements~\cite{chen2012antennas, muller2022angle}. 

Data-driven methods can be further divided into fingerprint-based~\cite{wang2017biloc, ni2022experience, hu2022experience} and learning-based approaches~\cite{arnold2019novel, ayyalasomayajula2020deep}. 
The former system build a dictionary mapping channel features to location ground truths offline, and during the online phase, it matches user device features to the closest key to provide the localization result. 
The latter system model the relationship between WiFi signals and coordinates as a black-box model using deep learning techniques. 

We belong to the learning-based approaches, but differentiate our work along four key axes. 
First, we introduce a pre-training phase before the conventional offline and online stages, which helps reduce the cost of fingerprint collection and enhances temporal robustness. 
Second, large-scale ISAC platforms present two challenges: heterogeneous inputs and missing values from some APs.
We address these issues by leveraging a novel graph structure to encode the heterogeneous inputs for the localization task, reducing redundancy in encoding and computation.
Third, unlike black-box models, we integrate a confidence-aware fine-tuning phase that improves localization performance and reduces the uncertainty inherent in black-box models. 
Finally, we evaluate our method on a large-scale server-side ISAC platform comprising over 40 or 400 APs.

\textbf{Graph Localization Techniques}:  
Graph Neural Networks (GNNs) have demonstrated exceptional capability in processing structured data across various domains, leading to their application in WiFi localization~\cite{zhou2020graph}.  
One class of graph-based approaches leverage the adjacency relationships between nodes and the information propagation mechanism to infer the labels of unlabeled nodes, which helps reduce the cost of constructing and maintaining offline fingerprint databases or models~\cite{wang2018improved, zheng2021grafin, chen2023graph, chiu2025graph}.
Several studies, similar to our approach, frame the localization task as a graph-level classification or regression problem~\cite{lezama2021indoor, sun2021novel, lezama2023application, wang2024graph, liu2025graph}. 
However, these methods primarily represent APs as nodes, utilizing homogeneous RSSI or CSI features as node attributes, or modeling location as a node with fixed-vector features. 
This approach overlooks both the heterogeneity of CSI dimensions and the issue of missing values, which are important in large-scale device deployments. 
Our work differs from these studies in two key aspects: first, our graph structure models the transmitter-receiver antenna pair as a node, with the corresponding CSI features of the link serving as the node attributes, which allows us to encode heterogeneous CSI.
Second, our graph structure does not encode APs that have not received WiFi signals, reducing redundancy in both encoding and computation.

\section{PROBLEM FORMULATION}
\label{section:formulation}

In this section, based on the wireless channel model in Sec.~\ref{subsection:channel_mode}, we formulate the localization problem in Sec.~\ref{subsection:loc_mode}. Then, we leverage two toy examples in Sec.~\ref{sec:wifi_signal_diversity} to demonstrate the heterogeneous nature of WiFi signals.

\begin{figure}[t]
	\centering
	\includegraphics[width=\columnwidth]{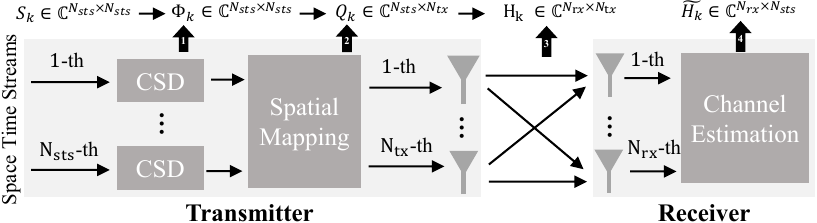}
	\caption{General structure of the wireless transmission process. The diagram illustrates the transmission process where spatial streams undergo CSD and spatial mapping at the transmitter. These streams then propagate through the physical channel environment and are received to calculate the CSI at the receiver end.}
	\label{fig:channelmodel}
\end{figure}

\subsection{Wireless Channel Model}
\label{subsection:channel_mode}
We begin by examining the general data transmission process in wireless communications, which consists of two main components: \textit{Training Sequence and Transmit Transformations}, and \textit{Channel Model}.
For clarity, the discussion focuses on a single transmitter-receiver pair.

\textit{Training Sequence and Transmit Transformations.}
Typical WiFi systems (e.g., IEEE 802.11n/ac/ax) use predefined training sequences at the receiver for precise channel estimation~\cite{tadayon2019decimeter, he2023sencom}. 
Formally, for each subcarrier \(k\), the transmit training matrix is defined as:
\begin{equation}
	\mathbf{S}_k \in \mathbb{C}^{N_{sts} \times N_{sts}},
\end{equation}
where \(N_{sts}\) denotes the number of space-time streams. Each row of this matrix corresponds to the training symbol vector for a respective stream~\cite{he2023sencom}. 

As illustrated in Fig.~\ref{fig:channelmodel}, the transmitter may apply transformations to optimize signal propagation.
First, space-time streams undergo cyclic shifting, represented by matrices:
\begin{equation}
	\mathbf{\Phi}_k \in \mathbb{C}^{N_{sts} \times N_{sts}}.
\end{equation}
This process, known as Cyclic Shift Diversity (CSD), improves frequency diversity and reduces unintended beamforming when transmitting identical information across multiple antennas~\cite{van2006802}.

Next, spatial mapping is applied through matrix:
\begin{equation}
	\mathbf{Q}_k \in \mathbb{C}^{N_{tx} \times N_{sts}},
\end{equation}
mapping \(N_{sts}\) space-time streams onto a greater number of transmit antennas, \(N_{tx}\). 
After these transformations, signals are emitted from the transmit antennas.

\textit{Channel Model.} 
The received signal at the receiver, denoted as \( \mathbf{Y}_k \), relates to the transmitted signal via the following matrix equation:
\begin{equation}
	\mathbf{Y}_k = \mathbf{H}_k \cdot \mathbf{Q}_k \cdot \mathbf{\Phi}_k \cdot \mathbf{S}_{k},
\end{equation}
where \( \mathbf{H}_k \in \mathbb{C}^{N_{rx} \times N_{tx}} \) represents the physical wireless channel, and \( N_{rx} \) is the number of receiving antennas.

The CSI estimated at the receiver, denoted as \(\widetilde{\mathbf{H}}_k \in \mathbb{C}^{N_{rx} \times N_{sts}}\), can thus be expressed as:
\begin{equation}
	\widetilde{\mathbf{H}}_k = \mathbf{Y}_k \cdot \mathbf{S}_k^{-1} = \mathbf{H}_k \cdot \left( \mathbf{Q}_k \cdot \mathbf{\Phi}_k \right),
\end{equation}
where \(\left(\mathbf{Q}_k \cdot \mathbf{\Phi}_k\right)\) represents transformations applied at the transmitter.

In scenarios prioritizing stability over throughput, dual-antenna devices commonly operate in single-stream mode (\(N_{sts}=1\))~\cite{tadayon2019decimeter, he2023sencom}, utilizing only one training symbol. 
In this mode, the training symbol undergoes both CSD and spatial mapping before being simultaneously transmitted from both antennas. The signals from these antennas combine into a single received signal, represented by \(\widetilde{\mathbf{H}}_k \in \mathbb{C}^{N_{rx} \times 1}\). 
Conversely, in multiplexing mode, each transmit antenna sends independent signals, resulting in distinct received signals represented as \(\widetilde{\mathbf{H}}_k \in \mathbb{C}^{N_{rx} \times 2}\).


Extending this concept, the CSI reported by the receivers, denoted as \(\widetilde{\mathbf{H}}\), can be represented as a three-dimensional tensor with dimensions \(N_{rx} \times N_{sts} \times N_{sub}\), where each of the \(N_{sub}\) matrices \(\widetilde{\mathbf{H}}_k\) corresponds to an individual subcarrier:
\begin{equation}
	\widetilde{\mathbf{H}} = [\widetilde{\mathbf{H}}_1, \widetilde{\mathbf{H}}_2, \dots, \widetilde{\mathbf{H}}_{N_{sub}}],
\end{equation}
where each \(\widetilde{\mathbf{H}}_k \in \mathbb{C}^{N_{rx} \times N_{sts}}\) denotes the CSI estimate for the \(k\)-th subcarrier.

For simplicity, we neglect the modeling of noise factors such as Carrier Frequency Offset (CFO), Sampling Frequency Offset (SFO), and others. For a more detailed treatment of these models, we refer the reader to~\cite{tadayon2019decimeter, zhang2019calibrating}.

\subsection{Problem Formulation for Localization}
\label{subsection:loc_mode}
We now extend the single-link channel model to a multi-receiver indoor localization scenario. 
Consider a set of user devices \(U\) (e.g., mobile phones, PCs, laptops) acting as transmitters, and a set of access points (APs) \(R\) deployed as receivers.

For a given transmitter \(u \in U\), we define a \textit{localization event} occurring within a short observation window \(\mathcal{W}_t\), typically one second in duration and centered at time \(t\). During this interval, multiple APs in the environment measure the CSI transmitted by device \(u\).
For clarity, we omit explicit details regarding the number of packets captured within this window \(\mathcal{W}_t\) and instead focus on aggregated CSI measurements. 

Let \(R^{(u, t)} \subseteq R\) be the subset of APs that collect CSI within the observation window:
\begin{equation}
	\label{eq:R_t}
	R^{(u, t)} = \left\{ r \in R \;\middle\vert\;
	\begin{array}{c}
		\text{AP } r \text{ measures CSI from} \\
		\text{transmitter } u \text{ during } \mathcal{W}_t
	\end{array} 
	\right\}.
\end{equation}
subject to the constraint \(|R^{(u,t)}| \ge 3\) to ensure sufficient spatial diversity for accurate localization~\cite{ni2022experience}.

Within this localization event, each AP \(r \in R^{(u,t)}\) captures one or more CSI matrices \(\widetilde{\mathbf{H}}^{(u,r,t,c)}\). 
In this context, the index \( c \) represents a specific transmission configuration, which includes several parameters, such as the transmission mode (e.g., diversity or multiplexing). 
Different transmission configurations may result in CSI matrices \(\widetilde{\mathbf{H}}^{(u,r,t,c)}\) having different dimensions due to variations in antennas, spatial streams, or subcarriers involved.
Formally, we denote the set of available transmission configurations at AP \( r \) as:
\[
\mathcal{C}^{(u,r,t)} = \left\{c_1, c_2, \dots, c_{|\mathcal{C}^{(u,r,t)}|}\right\},
\]
where each configuration \( c \in \mathcal{C}^{(u,r,t)} \) corresponds to a distinct set of parameters and consequently a distinct CSI matrix structure.

Formally, we define a localization event \(\mathbb{E}^{(u,t)}\) as the collection of all such CSI matrices across APs and supported transmission modes, paired with the ground-truth spatial coordinates \(\mathbf{l}^{(u,t)} \in \mathbb{R}^3\) of transmitter \(u\):
\begin{equation}
	\label{eq:locEvent}
	\mathbb{E}^{(u,t)} 
	= \Bigl\{
	\bigl\{\widetilde{\mathbf{H}}^{(u,r,t,c)}\bigr\}_{r \in R^{(u,t)},\, c \in \mathcal{C}^{(u,r,t)}},\,
	\mathbf{l}^{(u,t)}
	\Bigr\}.
\end{equation}

Our goal is to learn a predictive mapping \(f\) that takes the set of observed CSI matrices as input and outputs the corresponding spatial coordinates:
\begin{equation}
	\label{eq:mapping}
	f:\, \Bigl\{\widetilde{\mathbf{H}}^{(u,r,t,c)}\Bigr\}_{r \in R^{(u,t)},\, c \in \mathcal{C}^{(u,r,t)}}
	\;\longmapsto\; 
	\mathbf{l}^{(u,t)}.
\end{equation}

\subsection{Understanding Heterogeneous CSI}
\label{sec:wifi_signal_diversity}

\begin{figure}[h]
	\centering
	\begin{minipage}{0.5\columnwidth}  
		\centering
		\includegraphics[width=\textwidth]{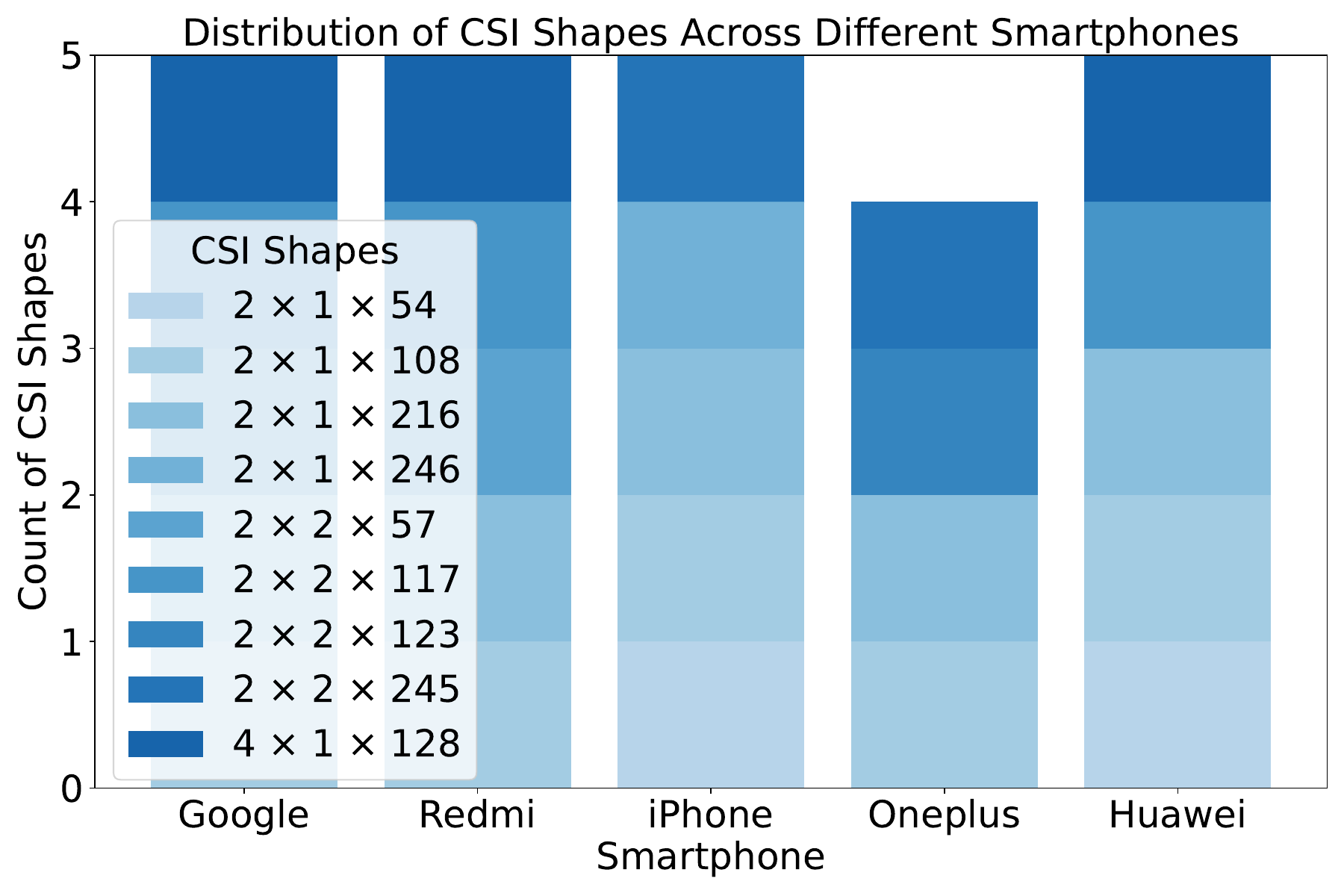}  
		\caption{The CSI dimensions of multiple users showcasing variations.}
		\label{fig:first_image}
	\end{minipage} \hfill  
	\begin{minipage}{0.45\columnwidth}  
		\centering
		\includegraphics[width=\textwidth]{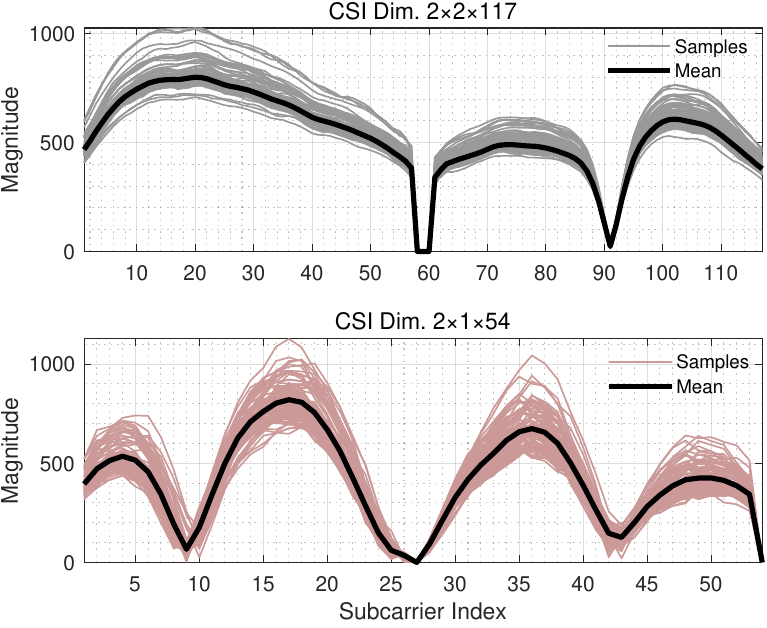}  
		\caption{Envelope plots of CSI amplitude captured at the same transmitter-receiver pair and location.}
		\label{fig:second_image}
	\end{minipage}
\end{figure}

To illustrate the heterogeneity of WiFi signals, we present two toy experiments that highlight this diversity from different perspectives. 

First, as illustrated in Fig.~\ref{fig:first_image}, we sample the CSI dimensions from multiple users and visualize their distribution patterns.
This experiment reveals the inherent structural differences in the CSI matrices.

Second, we collect packets from the same user device at the same location while communicating with the same AP. 
As shown in Fig.~\ref{fig:second_image}, for the same location, different CSI dimensions exhibit distinct envelope characteristics. 
In the multiplexing mode, one antenna link is selected for visualization. 
This suggests that simple aggregation strategies may be inadequate for handling the diversity across CSI dimensions, which is consistent with the findings in\cite{he2023sencom}.

\section{SYSTEM OVERVIEW}
\label{section:system}

\begin{figure}[t]
	\centering
	\includegraphics[width=\columnwidth]{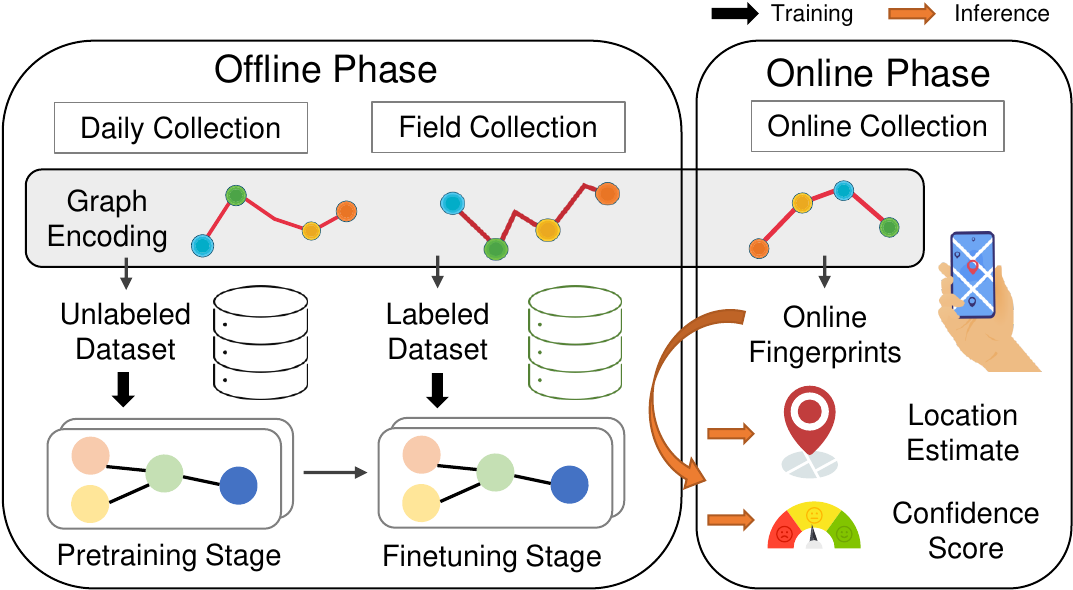}
	\caption{The processing pipeline of localization service of GLow.}
	\label{fig:framework}
\end{figure}

At a high level, as shown in Fig.~\ref{fig:framework}, the overall localization process is divided into two main stages: the offline and online phases. The offline phase consists of two key components: first, pretraining the model using continuously generated data; second, fine-tuning the model based on field-collected data. Subsequently, in the online phase, the system infers the location based on the fingerprints generated from signal measurements, producing both the localization result and its confidence estimation. 

In this section, we first introduce graph encoding for heterogeneous WiFi signals in Sec.~\ref{sec:graphencode}. Then, we present the system architecture and workflow in Sec.~\ref{sec:architecture}. In this part, we first describe the overall network architecture, followed by the spatiotemporal pretraining phase and the confidence-aware fine-tuning phase. Finally, we explain how the online phase estimates the location and confidence level.

\subsection{Graph Encoding of Input for Localization Events}
\label{sec:graphencode}
In this part, we detail our approach for encoding the diverse WiFi signals for neural network input, as previously described, within a localization event \(\mathbb{E}^{(u,t)}\).
Traditional systems represent these signals as fixed-length vectors or matrices, where each row corresponds to a different AP, and each column represents a WiFi signal sample.
However, they may suffer from the missing-value problem~\cite{zhuo2022grafics}, as not all signals from the APs at a given location are fully scanned, leaving several entries in the matrix empty.
These missing entries are typically filled with arbitrarily small values, which can introduce unintended artifacts in the feature learning process~\cite{chiu2025graph}.
Not all the signals from the APs at a given location may be fully scanned, leaving several entries in the vectors or matrices empty. These missing entries are typically filled with arbitrarily small values, which may introduce unintended artifacts in the feature learning process.

To address these challenges, we propose constructing a graph representation \(\mathcal{G}^{u,t}\) to encode the heterogeneous CSI in a localization event \(\mathbb{E}^{u,t}\).
We begin by constructing the graph \(\mathcal{G}^{u,t} = (\mathcal{V}, \mathcal{E})\), where \(\mathcal{V}\) denotes the set of nodes and \(\mathcal{E}\) represents the set of edges.
In this context, the nodes correspond to the signal features from different receiver antennas, including CSI magnitudes~\cite{wang2016csi}, Channel Impulse Response (CIR) magnitudes~\cite{wu2012fila}, and phase differences between signals from different receiver antennas~\cite{wang2016csi2}. The edges capture the relationships between these features. 

\textbf{Node Definition:}
Each node \(v_i \in \mathcal{V}\) represents a specific feature extracted from the CSI matrices, which are associated with the signal samples received from different antennas of various APs or from the interactions between antennas of the same AP.
Specifically, we define the features for each node as:

\[
\mathbf{x}_i = \text{A feature vector from } \widetilde{\mathbf{H}}^{(u,r,t,c)} \in \mathbb{E}^{(u,t)},
\]

where each feature is derived from the CSI and CIR magnitudes of each transmit-receive link, as well as the phase differences between signals received from different antennas within the same transmit spatial stream. 
Formally, the CSI matrix \(\widetilde{\mathbf{H}}^{(u,r,t,c)}\) is a complex-valued matrix, and the magnitude of the CSI at each spatial stream \(n_{\text{sts}}\) and receiver antenna pair \(n_{\text{rx}}\) is computed as:

\begin{equation}
	|\widetilde{\mathbf{H}}^{(u,r,t,c)}_{[n_{\text{rx}},n_{\text{sts}}, :]}| = \sqrt{\Re\left(\widetilde{\mathbf{H}}^{(u,r,t,c)}_{[n_{\text{rx}},n_{\text{sts}}, :]}\right)^2 + \Im\left(\widetilde{\mathbf{H}}^{(u,r,t,c)}_{[n_{\text{rx}},n_{\text{sts}}, :]}\right)^2}
\end{equation}

where \(\Re\left(\widetilde{\mathbf{H}}^{(u,r,t,c)}_{[n_{\text{rx}},n_{\text{sts}}, :]}\right)\) and \(\Im\left(\widetilde{\mathbf{H}}^{(u,r,t,c)}_{[n_{\text{rx}},n_{\text{sts}}, :]}\right)\) represent the real and imaginary parts of the CSI matrix, respectively.
The CIR magnitudes is derived from the CSI matrix through an inverse Fourier transform, capturing the multipath characteristics of the channel:

\begin{equation}
		|\text{CIR}_{\text{time}}^{(u,t,t,c)}| = \left|\mathcal{F}^{-1}\left(\widetilde{\mathbf{H}}^{(u,r,t,c)}\right)\right| \\
\end{equation}

where \(\mathcal{F}^{-1}\) denotes the inverse Fourier transform operator. Additionally, phase differences between signals received at different antennas are essential for understanding relative signal delays.
For each spatial stream \(n_{\text{sts}}\), the phase difference between two receiver antennas \(n_{\text{rx}_i}\) and \(n_{\text{rx}_j}\) is computed as:
\begin{equation}
	\Delta\theta_{n_{\text{rx}_i}, n_{\text{rx}_j}}^{(u,r,t,c)} = \arg\left(\widetilde{\mathbf{H}}^{(u,r,t,c)}_{[n_{\text{rx}_i},n_{\text{sts}}, :]} - \widetilde{\mathbf{H}}^{(u,r,t,c)}_{[n_{\text{rx}_j},n_{\text{sts}}, :]}\right)
\end{equation}
where \(\arg(\cdot)\) returns the phase of the complex-valued CSI entries. 
More specifically, feature vectors are standardized to a fixed dimension of 245, with zero-padding applied where necessary and downsampling performed when the dimension exceeds this threshold.
Additionally, the position of AP, the central frequency of the CSI, and the RSSI values are recorded and incorporated as part of the node features for each node. 

\begin{figure}[b]
	\centering
	\includegraphics[width=0.6\columnwidth]{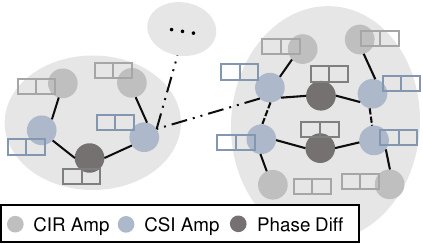}
	\caption{Definition of Edge Connections.}
	\label{fig:csiencode}
\end{figure}

\textbf{Edge Definition:} The edges \(\mathcal{E}\) capture relationships between nodes. Specifically, we establish edges based on the following criteria, as shown in Fig.~\ref{fig:csiencode}:

\begin{itemize}
	\item CSI amplitude nodes corresponding to different receiver antennas are connected through phase difference nodes in each \(\mathbf{H}^{(u,r,t,c)}\).
	\item CSI and CIR amplitude nodes within the same receiver antenna are connected in each \(\mathbf{H}^{(u,r,t,c)}\).
	\item CSI amplitude nodes originating from the same spatial stream are interconnected in each \(\mathbf{H}^{(u,r,t,c)}\).
	\item CSI amplitude nodes with the highest RSSI in each \(\mathbf{H}^{(u,r,t,c)}\) act as bridges, connecting to nodes in other \(\mathbf{H}^{(u,r,t,c)}\) instances.
\end{itemize}
The adjacency matrix \(A \in \mathbb{R}^{N_v \times N_v}\) represents the connectivity between nodes, where:

\[
A_{ij} = \begin{cases} 
	1, & \text{if there is an edge between nodes } v_i \text{ and } v_j \\
	0, & \text{otherwise}
\end{cases}
\]

This completes the construction of the graph representation \(\mathcal{G}^{u,t}\) for the event \( \mathbb{E}^{(u,r,t,c)} \).

\subsection{System Architecture and Workflow}
\label{sec:architecture}

\subsubsection{Network Architecture}
\begin{figure}[t]
	\centering
	\includegraphics[width=\columnwidth]{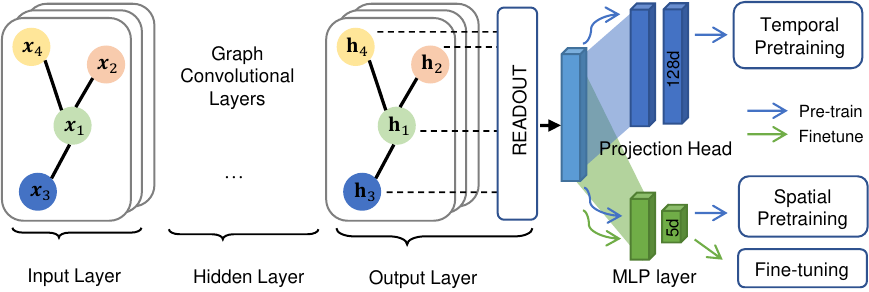}
	\caption{The Network Architecture of GLow.}
	\label{fig:network_framework}
\end{figure}

Fig.~\ref{fig:network_framework} illustrates the three main components of GLow’s neural network architecture:
(1) a graph-based neural network encoder, 
(2) a projection head, and 
(3) a Multi-Layer Perceptron (MLP). 
During pretraining, these components are jointly optimized to learn comprehensive spatiotemporal representations. In the fine-tuning phase, however, only the GNN encoder and MLP are updated, while the projection head remains fixed.

\textit{Graph-Based Neural Network Encoder}. 
We employ an iterative neighborhood aggregation (also known as message passing) scheme~\cite{kipf2016semi,you2020graph} to capture the structural information within the nodes' neighborhoods.  
Formally, let \( \mathcal{G} = (\mathcal{V}, \mathcal{E}) \) denote the input graph for a localization event, where \( \mathcal{V} \) is the set of nodes and \( \mathcal{E} \) is the set of edges. 
The feature matrix is represented by \( \mathbf{X} \in \mathbb{R}^{|\mathcal{V}| \times N_{node}} \), with \( \mathbf{x}_n = \mathbf{X}[n, :]^T \) as the \( N_{node} \)-dimensional attribute vector of node \( v_n \).

Specifically, we use the Graph Convolutional Network (GCN)~\cite{kipf2016semi} as the encoder backbone, with a default depth of \( L = 4 \) layers. 
In an \(L\)-layer graph neural network (GNN) encoder \(g(\cdot)\), the embedding propagation at the \(l\)-th layer is computed by:
\begin{align}
	\mathbf{a}_n^{(l)} &= \text{AGGREGATE}^{(l)}(\{ \mathbf{h}_{n'}^{(l-1)} : n' \in \mathcal{N}(n) \}), \\
	\mathbf{h}_n^{(l)} &= \text{COMBINE}^{(l)}(\mathbf{h}_n^{(l-1)}, \mathbf{a}_n^{(l)}),
\end{align}
where \( \mathbf{h}_n^{(l)} \) represents the embedding of node \( v_n \) at layer \( l \) with \( h_n^{(0)} = \mathbf{x}_n \).
Here, \( \mathcal{N}(n) \) represents the set of nodes adjacent to \( v_n \), and \( \text{AGGREGATE}^{(l)}(\cdot) \) and \( \text{COMBINE}^{(l)}(\cdot) \) function as component functions of the GNN layer.

After \(L\) layers of propagation, the final graph-level embedding \(g(\mathcal{G})\) is obtained by synthesizing node embeddings through a READOUT function:
\begin{equation}
	g(\mathcal{G}) = \text{READOUT}(\{ \mathbf{h}_n^{(l)} : v_n \in \mathcal{V}, l \in L \}).
\end{equation}

\textit{Projection Head and MLP}. 
From \( g(\mathcal{G}) \), two separate modules produce different outputs. 
A projection head (a two-layer perceptron) generates a 128-dimensional vector \( \mathbf{z} \)~\cite{chen2020simple} for temporal pretraining, while an MLP yields a vector \( \mathbf{d} \) for spatial pretraining. 
\begin{align}
	\mathbf{z} &= \text{PROJECTION HEAD}(g(\mathcal{G})), \\
	\mathbf{d} &= \text{MLP}(g(\mathcal{G})).
\end{align}

The MLP consists of three layers, with each intermediate layer reducing its neuron count by half compared to the preceding one, except for the final layer which outputs \( \mathbf{d} \). The \text{ReLU} activation function is applied throughout both the projection head and the MLP, except at their final output layers.

\begin{figure}[t]
	\centering
	\includegraphics[width=1.0\columnwidth]{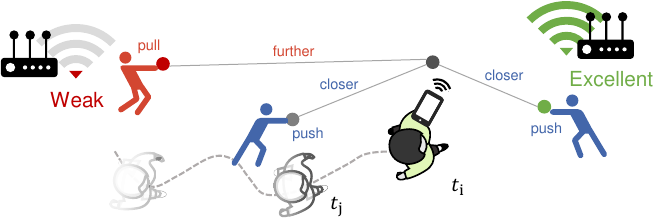}
	\caption{The spatiotemporal prior demonstrates that CSI samples collected from the same user within a short time interval are typically associated with nearby locations in space, while the spatial component of this prior suggests that APs with higher received signal power are generally closer to the user device.}
	\label{fig:spatiotemporal}
\end{figure}

\subsubsection{Spatiotemporal Pretraining Phase}
\label{pretraining}

In the ISAC system, large-scale unlabeled CSI data is readily available during its operation. 
Although these data lack explicit positional annotations, they implicitly contain valuable spatiotemporal prior information.
Specifically, two key priors can be leveraged:

\textbf{Temporal Prior:} User movement is inherently constrained by physical mobility patterns, resulting in spatially continuous data when CSI samples are taken closely in time.
As illustrated in Fig.~\ref{fig:spatiotemporal}, samples captured within brief intervals naturally represent similar spatial contexts, while increasing time gaps or sampling from distinct users significantly reduces spatial correlation.
For server-side ISAC deployments, timestamps are readily accessible without additional cost, making temporal information an efficient means to implicitly encode spatial continuity into model training. 

\textbf{Spatial Prior:} WiFi signal propagation inherently encodes spatial information. 
As depicted in Fig.~\ref{fig:spatiotemporal}, the RSSI measured across multiple APs directly correlates with their proximity to the transmitting device.
APs with higher RSSI values are typically closer, particularly under line-of-sight conditions. 
Furthermore, the positions of APs\cite{taner2023channel} are commonly used in localization systems.
From a server-side perspective, acquiring this information is relatively straightforward~\cite{ni2022experience, taner2023channel}, often through CAD layouts or site surveys. 
This spatial prior introduces broader structural constraints, guiding the model towards representations consistent with the underlying spatial topology. 

Motivated by these insights, we propose a unified pretraining framework that simultaneously exploits both temporal and spatial priors.
The primary objective of this pretraining is to learn robust spatiotemporal representations that enhance downstream localization performance. 
Specifically, we optimize the GNN encoder by jointly leveraging two pretext tasks:

\begin{algorithm}[t]
\caption{PyTorch-style Pretraining Pseudocode}
\label{alg:pretraining}
\begin{lstlisting}
# Pretraining Phase (B: batch size; F: feature dim; D/D': projection dims)
# Load a batch of anchor-positive pairs
for G_anc, G_pos in loader:
	# Feature encoding (B x F)
	feat_anc, feat_pos = g(G_anc), g(G_pos)
	# Temporal embedding projection (B x D)
	z_anc, z_pos = PROJECTION_HEAD(feat_anc), PROJECTION_HEAD(feat_pos)
	# Spatial embedding projection (B x D')
	d_anc, d_pos = MLP(feat_anc), MLP(feat_pos)
	
	# Temporal contrastive loss
	L_t = NT_XentLoss(z_anc, z_pos, temperature)
	# Spatial & Floor losses (metric + supervision)
	L_s = TripletLoss(d_anc[:, :3], r_p_anc, r_n_anc) + TripletLoss(d_pos[:, :3], r_p_pos, r_n_pos)
	L_f = L1Loss(d_anc[:,2], r_p_anc[:,2]) + L1Loss(d_pos[:,2], r_p_pos[:,2])
	# Total pretraining loss
	L_pretrain = L_t + L_s + L_f
	# ..
\end{lstlisting}
\end{algorithm}

\textbf{Temporal Pretraining:} We utilize a SimCLR-based framework adapted for graph data. Given a graph \(\mathcal{G}^{(u,t_i)}\) from user device \(u\) at timestamp \(t_i\), we consider another graph  \(\mathcal{G}^{(u,t_j)}\) from the same user as a positive sample if their timestamps satisfy \(|t_i - t_j| \leq 6\) seconds\footnote{This threshold is chosen based on the specific system sampling rates and the observed mobility patterns of users.}.
The embeddings of these graph pairs are encouraged to be similar using the NT-Xent loss~\cite{chen2020simple}, which is defined as:
\begin{equation}
	\mathrm{Num} = \exp\Bigl(\text{sim}\bigl(\mathbf{z}^{(u,t_i)}, \mathbf{z}^{(u,t_j)}\bigr)/\tau\Bigr),
\end{equation}

\begin{equation}
	\mathrm{Den} = \sum_{k=1}^{2N_{bs}} \mathbf{1}_{[k \neq (u,t_i)]} \exp\Bigl(\text{sim}\bigl(\mathbf{z}^{(u,t_i)}, \mathbf{z}^{k}\bigr)/\tau\Bigr),
\end{equation}

\begin{equation}
	\mathcal{L}_t\Bigl(\mathcal{G}^{(u,t_i)}, \mathcal{G}^{(u,t_j)}\Bigr)
	= -\log \left(\frac{\mathrm{Num}}{\mathrm{Den}}\right).
\end{equation}

Here, $\mathbf{z}^{(u,t)}$ denotes the embedding of the graph $\mathcal{G}^{(u,t)}$, $\tau$ is the temperature parameter,  which is empirically set to 0.5.
$\text{sim}(\cdot, \cdot)$ denotes the cosine similarity function defined as

\begin{equation}
	\text{sim}\Bigl(\mathbf{z}^{(u,t_i)}, \mathbf{z}^{(u,t_j)}\Bigr) = \frac{\Bigl(\mathbf{z}^{(u,t_i)}\Bigr)^\top \mathbf{z}^{(u,t_j)}}{\|\mathbf{z}^{(u,t_i)}\| \, \|\mathbf{z}^{(u,t_j)}\|}.
\end{equation}


\textbf{Spatial Pretraining:} To effectively encode spatial priors, we employ two learning strategies. 

First, we apply a triplet loss~\cite{taner2023channel} to enforce geometric consistency in the learned embeddings.
For each sample, the anchor is predicted from input features, the positive is defined as the AP with the strongest RSSI, and negatives are selected from APs with 5–20 dBm weaker signals. The objective encourages embeddings to reflect relative spatial layout:

\begin{equation}
	\mathcal{L}_{s} (\mathcal{G}^{(u,t)}) = \max\left(0, \left\| \mathbf{d}_{[1:3]}^{(u,t)} - \mathbf{r}_p^{(u,t)} \right\|_2^2 - \left\| \mathbf{d}_{[1:3]}^{(u,t)} - \mathbf{r}_n^{(u,t)} \right\|_2^2 + \alpha \right),
\end{equation}


where \(\mathbf{d} \in \mathbb{R}^5\) is the predicted embedding vector, and \(\mathbf{d}_{[1:3]} \in \mathbb{R}^3\) corresponds to the estimated 3D spatial position, while the ground-truth location is denoted as \(\mathbf{l}^{(u,t)}\). 
\(\mathbf{r}_p^{(u,t)}\) and \(\mathbf{r}_n^{(u,t)}\) represent the physical positions of the positive and negative APs associated with sample \((u,t)\), respectively, and \(\alpha\) is a margin hyperparameter empirically set to 3.


Second, we incorporate vertical supervision by regressing the floor level of the strongest AP. This encourages the model to preserve inter-floor differences:

\begin{equation}
	\mathcal{L}_f (\mathcal{G}^{(u,t)}) = \left| \mathbf{d}^{(u,t)}_{[3]} - \mathbf{r}^{(u,t)}_{p,[3]} \right|,
\end{equation}
where \(\mathbf{d}^{(u,t)}_{[3]}\) represents the predicted floor level for sample \((u,t)\), and \(\mathbf{r}^{(u,t)}_{p,[3]}\) is the corresponding floor level of the strongest AP. 


Ultimately, these losses are jointly optimized through a weighted combination to form the overall pretraining objective:
\begin{equation}
	\mathcal{L}_{\text{pretrain}} = \mathcal{L}_t + \mathcal{L}_s + \mathcal{L}_f,
\end{equation}
while implementation details of the pretraining phase are illustrated in the PyTorch-style pseudocode shown in Algorithm~\ref{alg:pretraining}.

%
%

\subsubsection{Confidence-Aware Fine-Tuning Phase}
\label{fintuning}

In this section, we introduce a confidence-aware fine-tuning strategy that explicitly accounts for prediction uncertainty. 
Specifically, our model not only outputs the \(\mathbf{d}_{[1:3]}^{(u,t)}\), where the corresponding ground truth \(\mathbf{l}^{(u,t)} \in \mathbb{R}^3\), but also provides a confidence score \(\mathbf{d}_{[4:5]}^{(u,t)}\) for each predicted coordinate. 
We omit the confidence score for floor prediction, as its accuracy is already sufficiently high. 

In the fine-tuning training phase, we model the ground truth \(\mathbf{l}_{[1:2]}^{(u,t)}\) as two separate Dirac delta functions \(\delta_{f_1}\) and \(\delta_{f_2}\), each centered at the corresponding target coordinates in the 2D space. 
Next, we model the predicted probabilistic distributions of the coordinates for each dimension of the target \(\mathbf{l}_{[1]}^{(u,t)}\) and \(\mathbf{l}_{[2]}^{(u,t)}\) by treating \(\mathbf{d}_{[1]}^{(u,t)}\), \(\mathbf{d}_{[4]}^{(u,t)}\), \(\mathbf{d}_{[2]}^{(u,t)}\), and \(\mathbf{d}_{[5]}^{(u,t)}\) as the model’s predicted probabilistic estimates. These predicted distributions are modeled as Gaussian distributions \(\mathcal{N}(\mathbf{d}_{[1]}^{(u,t)}, \mathbf{d}_{[4]}^{(u,t)})\) and \(\mathcal{N}(\mathbf{d}_{[2]}^{(u,t)}, \mathbf{d}_{[5]}^{(u,t)})\), corresponding to each dimension of the target coordinates. 

To train the model, we minimize the KL Divergence between the predicted probabilistic distributions and the Dirac delta functions for each dimension, using the following general formulation:

\begin{align}
	\mathcal{L}_{\text{KL}}(\mathbf{l}_{[i]}^{(u,t)}, \mathbf{d}_{[i,4+i]}^{(u,t)}) &= D_{\text{KL}} \left( \mathcal{N}(\mathbf{d}_{[i]}^{(u,t)}, \mathbf{d}_{[4+i]}^{(u,t)}) \parallel \delta_{f_i} \right) \nonumber \\
	&\propto \frac{\log(\mathbf{d}_{[4+i]}^{(u,t)})}{2} + \frac{\left( \mathbf{l}_{[i]}^{(u,t)} - \mathbf{d}_{[i]}^{(u,t)} \right)^2}{2 \left( \mathbf{d}_{[4+i]}^{(u,t)} \right)^2} \\
	&\quad \quad \quad \text{for } i = 1, 2.
\end{align}

Finally, the total loss function is the sum of the two KL divergence terms and the L1 loss for \(i = 3\):
\begin{equation}
	\mathcal{L}_{\text{fine-tuning}} = \sum_{i=1}^{2} \mathcal{L}_{\text{KL}}(\mathbf{l}_{[i]}^{(u,t)}, \mathbf{d}_{[i,4+i]}^{(u,t)}) + \mathcal{L}_{L1}(\mathbf{l}_{[3]}^{(u,t)}, \mathbf{d}_{[3]}^{(u,t)}),
\end{equation}
where the L1 loss \(\mathcal{L}_{L1}\) is defined as:
\begin{equation}
	\mathcal{L}_{L1}(\mathbf{l}_{[3]}^{(u,t)}, \mathbf{d}_{[3]}^{(u,t)}) = \left\| \mathbf{l}_{[3]}^{(u,t)} - \mathbf{d}_{[3]}^{(u,t)} \right\|_1.
\end{equation}

Based on the loss function described above, in the fine-tuning training phase, we train $Z=5$ models independently, each with the same structure and training data but different random seeds. 

\subsubsection{Online Phase}
In the online phase (or test phase), we compute the final estimation of the coordinates and their associated confidence scores using the following weighted average logic.
The final predicted coordinates \(\tilde{\mathbf{d}}_{[i]}^{(u,t)}\) for each dimension \(i\) are computed as follows:

\begin{equation}
	\tilde{\mathbf{d}}_{[i]}^{(u,t)} = \frac{\sum_{z=1}^{Z} \left( \frac{\mathbf{d}_{[i]}^{(u,t)}}{\mathbf{d}_{[4+i]}^{(u,t)}} \right)}{\sum_{z=1}^{Z} \left( \frac{1}{\mathbf{d}_{[4+i]}^{(u,t)}} \right)}, \quad \text{for } i = 1, 2.
\end{equation}

where \(\mathbf{d}_{[i]}^{(u,t)}\) represents the predicted coordinate from the \(z\)-th model, and \(\mathbf{d}_{[4+i]}^{(u,t)}\) represents its associated confidence score.

For the floor prediction, we compute the average of the floor estimates from all models:

\begin{equation}
	\tilde{\mathbf{d}}_{[3]}^{(u,t)} = \frac{\sum_{z=1}^{Z} \mathbf{d}_{[3]}^{(u,t)}}{Z},
\end{equation}

where \(\mathbf{d}_{[3]}^{(u,t)}\) represents the predicted floor level from the \(z\)-th model. We then round \(\mathbf{d}_{[z]}\) to the nearest integer to determine the final floor level.

For the confidence score, we compute it as the average of the confidence scores from all models:

\begin{equation}
	\tilde{\mathbf{d}}_{[j]}^{(u,t)} = \frac{\sum_{z=1}^{Z} \mathbf{d}_{[j]}^{(u,t)}}{Z}, \quad \text{for } j = 4, 5.
\end{equation}

where \(\mathbf{d}_{[j]}^{(u,t)}\) represents the confidence score for the predicted coordinates, and \(\tilde{\mathbf{d}}_{[j]}^{(u,t)}\) is the final averaged confidence score.

\section{Dataset Description}
\label{section:dataset}
In this section, we first describe the unlabeled dataset, followed by the single-floor and multi-floor datasets.

\subsection{Unlabelled Dataset}
We collected data accumulated during daily operations over a 29-day period, resulting in 85 GB of raw data across five floors of the building.  
To mitigate data redundancy, we applied downsampling based on RSSI similarity~\cite{wu2012will}.  
From the downsampled data, 120k unlabeled samples were selected to construct the pretraining set.  
These samples are entirely disjoint from those used in subsequent experimental evaluations.

\subsection{Single-Floor Dataset}

\begin{figure}[t]
	\centering
	\includegraphics[width=1.0\columnwidth]{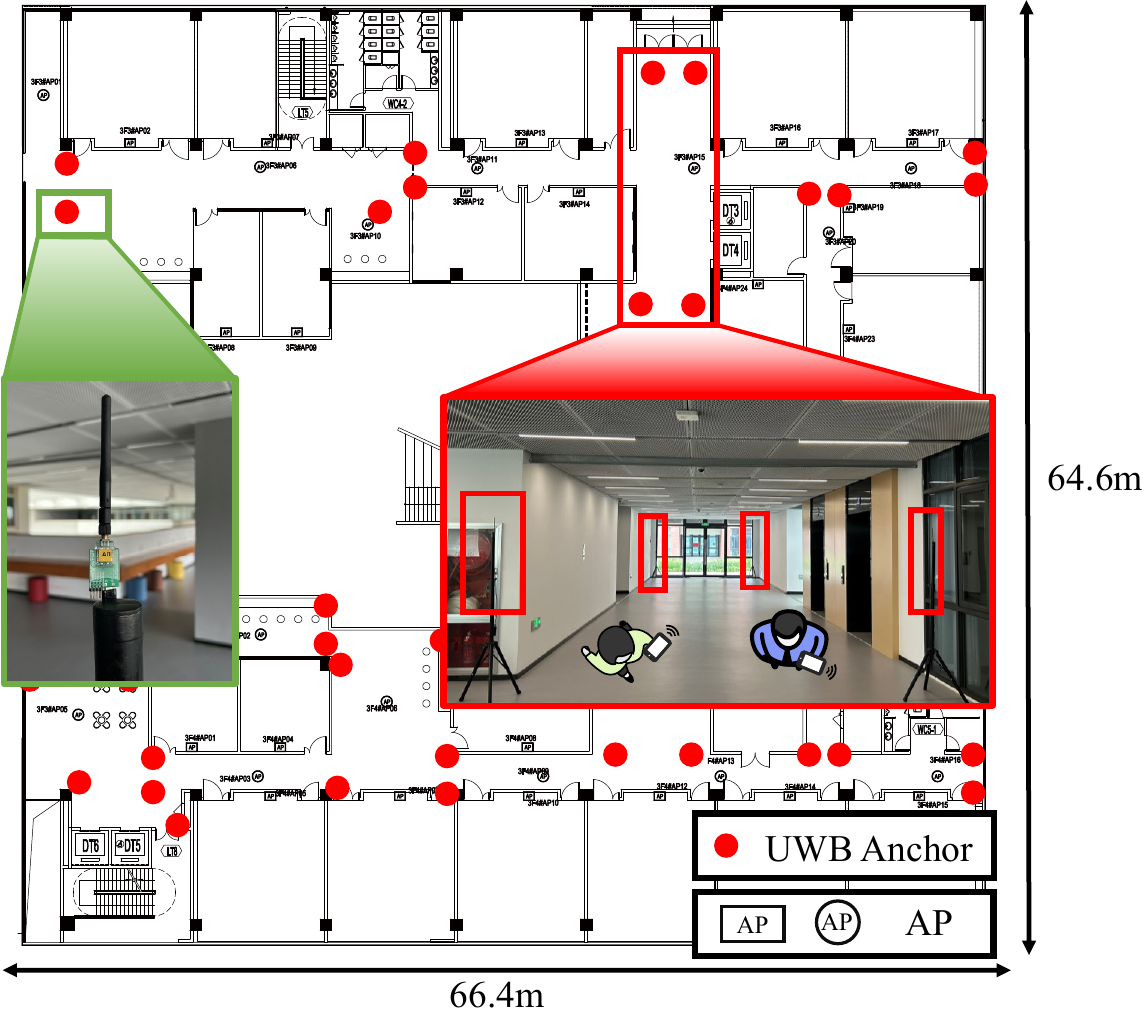}
	\caption{Floor-level experimental setup. Red dots indicate the positions of UWB base stations, while APs are denoted by square and circular markers. The red box highlights one of the sections where data collection was conducted.}
	\label{fig:uwb_2dfloorplan}
\end{figure}

We conduct our evaluations in an indoor environment spanning approximately \(4,000 \, \text{m}^2\), as illustrated in Fig.~\ref{fig:uwb_2dfloorplan}.  
Ground-truth labels are obtained using an Ultra-Wideband (UWB) localization system~\cite{zhang2024rloc}.  
The experimental area is segmented into smaller sections to ensure that each maintains Line-of-Sight (LoS) conditions suitable for UWB deployment.  
In each section, four UWB anchors are mounted on tripods at a uniform height of 1.8 meters, guaranteeing unobstructed LoS above potential obstacles.  
This configuration provides precise ground-truth labels with a localization accuracy of up to ten centimeters~\cite{zhang2024rloc}.  
A battery-powered UWB tag is affixed to the user's head via a custom-designed helmet, enabling natural movements such as walking at varying speeds and pausing, effectively simulating daily human activities.  
A Network Time Protocol (NTP) server is deployed to synchronize all devices with millisecond-level precision.  
Importantly, during data collection, smartphones are either placed in pockets or held in hand to closely reflect real-world usage scenarios.  
Our data collection emphasizes five distinct smartphone models: iPhone 14, Huawei P30, Honor X10, Redmi K20 Pro, and Pixel 4.  
This dataset, consisting of approximately 30,000 samples, serves as the foundation for our training, validation, and testing stages, divided in an 8:1:1 ratio.


\subsection{Multi-Floor Leave-one-phone-out Dataset}
\begin{figure}[t]
	\centering
	\includegraphics[width=1.0\columnwidth]{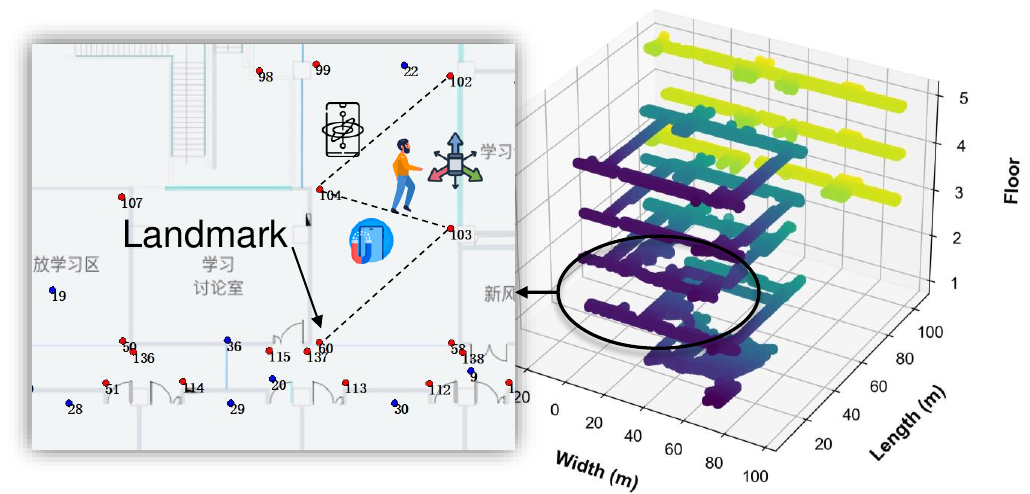}
	\caption{Multi-Floor Dataset. The right side shows the 3D distribution of the collected fingerprint points, while the left image illustrates the landmark-based collection process.}
	\label{fig:waypoint}
\end{figure}

To further assess the performance of the GLow prototype in more complex and large-scale environments, we expand both the scope and intricacy of the dataset. 
On one hand, we collect approximately 40,000 samples from seven distinct devices: iPhone 14, Huawei P30, Honor X10, Redmi K20 Pro, Pixel 4, Huawei Mate10, and Mi8SE. These samples span approximately \(25,600 \, \text{m}^2\) across five floors, as shown in Fig.~\ref{fig:waypoint}. 
We adopt a fingerprint collection scheme based on landmarks and pedestrian trajectory tracking, as detailed in \cite{hu2022experience, hu2023wisdom}. 
On the other hand, we evaluate the leave-one-phone-out performance, where a specific mobile phone model is excluded from the training dataset and used exclusively for testing. 
This evaluation setting simulates realistic usage scenarios, in which mobile devices encountered during daily location-based services may differ from those present in the training set.

\section{EXPERIMENTAL EVALUATION}
\label{section:evaluation}
In this section, we first introduce the baseline algorithms, evaluation metrics, and implementation details, followed by an analysis of the overall performance in Sec.\ref{overall} and a detailed evaluation of performance experiments in Sec.\ref{detail}.

\textbf{Baseline algorithms}: We compare our system with baseline localization methods, including both traditional and learning-based approaches. Below we briefly describe the representative baselines used in our evaluation:

\begin{itemize}
	
	\item \textbf{k-NN}~\cite{torres2015comprehensive}: A classical fingerprinting method that estimates the user's location by computing the distance between the current RSSI vector and those in the labeled dataset.
	
	\item \textbf{CNNLoc}~\cite{jang2018indoor}: A deep learning-based method that leverages convolutional neural networks to extract spatial features from RSSI inputs for indoor localization.
	
	\item \textbf{DLM}~\cite{arnold2019novel}: A deep learning baseline model for CSI-based localization that uses two convolutional layers, followed by average pooling and fully connected layers.
	
	\item \textbf{MSG}~\cite{liu2025graph}: A graph neural network-based method that utilizes multiscale principal component analysis to process features and constructs a CSI graph for localization. 
	
	\item \textbf{CNNLoc+}~\cite{sandler2018mobilenetv2, jang2018indoor}: We extend the approach in~\cite{jang2018indoor} to incorporate CSI feature inputs. The network architecture uses~\cite{sandler2018mobilenetv2} as its backbone.
	
\end{itemize}

It should be noted that, because the above system does not consider the heterogeneity of the WiFi signal, it necessitates omitting certain data to maintain uniform dimensions across the dataset, as previously discussed. 
Some methods~\cite{ayyalasomayajula2020deep, zhang2024rloc} convert CSI signals into two-dimensional likelihood heatmaps that represent angle-of-arrival and time-of-flight (or distance) information. 
However, when scaling to systems with over 40 or even 400 APs, such input formats incur a substantial memory overhead, which may negatively impact the system's inference latency.

To evaluate the performance of pretraining strategies, we compare with state-of-the-art methods:
\begin{itemize}
	\item \textbf{GCL}~\cite{you2020graph}: A contrastive learning approach that employs four graph augmentations: node dropping, edge perturbation, attribute masking, and subgraph creation, to generate positive sample pairs.
	
	\item \textbf{Tri.}~\cite{ferrand2021triplet}: A triplet-based learning method that maps CSI samples into a low-dimensional space, leveraging prior knowledge from the time dimension. 
	
	\item \textbf{Pseudo.}~\cite{ni2022experience}: A method that utilizes coarse location estimates derived from AP positions and RSSI values as pseudo-labels for fingerprint initialization. 
	
	\item \textbf{GMAE}~\cite{hou2022graphmae}: A generative method that performs representation learning through feature reconstruction from masked inputs, using a masking strategy with a consistent mask rate and scaled cosine error.	
	
	\item \textbf{Tri.+Bi.}~\cite{taner2023channel}. A triplet-based learning method that maps CSI samples into a low-dimensional space, leveraging prior knowledge from both the time and spatial dimensions. 
	
\end{itemize}

\textbf{Evaluation metrics:}
To measure the accuracy of location prediction, model performance is assessed using a combined score that incorporates both positioning error and floor estimation error: $\text{score} = \sqrt{(x - \hat{x})^2 + (y - \hat{y})^2} + p \cdot |\hat{f} - f|$, where \( p \) is the penalty factor, set to $15$, following the methodology in \cite{hu2022experience}. 

\textbf{Implementation details}: Our experimental framework comprises three distinct phases: pre-training, training from random initialization, and fine-tuning.
In the pre-training phase, models are pre-trained for 150 epochs with a learning rate of 0.01 and a batch size of 4096 using the Adam~\cite{kingma2014adam} optimizer with default parameters. 
Unless otherwise specified, the margin \(\alpha\) and temperature $\tau$ are set to 3 and 0.5 by default.
For models trained from scratch, we use a learning rate of 0.001, a batch size of 64, and train for 100 epochs. 
The \texttt{ReduceLROnPlateau} scheduler adjusts the learning rate based on validation performance.
Fine-tuning lasts 100 epochs with different learning rates applied to different parts of the network: the first three layers of the GCN are fine-tuned with a learning rate of 0.001, while the remaining GCN layers and MLP are fine-tuned with 0.01. 
The batch size remains 64, and the same optimizer and scheduler as used in training from scratch are employed. 

\subsection{Overall Performance Experiment}
\label{overall}

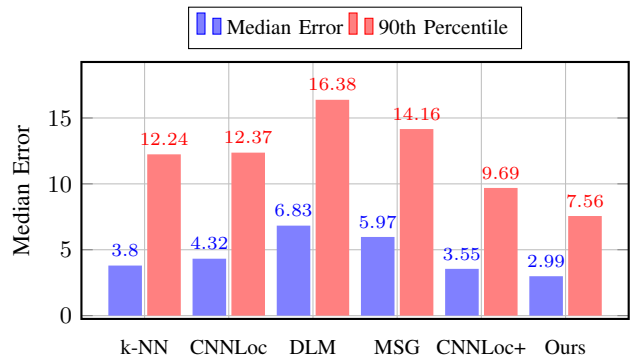
\begin{figure}[t]
	\centering
	\begin{tikzpicture}
		\begin{axis}[	
			ybar=2pt,
			bar width=12.7pt,
			width=1.0\columnwidth,
			height=5cm,
			ymin=2, ymax=17,
			ylabel={Median Error},
			symbolic x coords={k-NN, CNNLoc, DLM, MSG, CNNLoc+, Ours},
			xtick=data,
			xtick style={draw=none},
			ylabel style={font=\small},
			tick label style={font=\small},
			legend style={
				at={(0.5,1.05)},
				anchor=south,
				legend columns=2,
				font=\footnotesize
			},
			nodes near coords,
			nodes near coords align={vertical},
			every node near coord/.append style={font=\scriptsize},
			axis line style=thick,
			grid=major,
			enlargelimits=0.15,
			xticklabel style={font=\footnotesize},  
			]
			
			\addplot+[draw=none, fill=blue!50] coordinates {
				(k-NN, 3.80) (CNNLoc, 4.32) (DLM, 6.83) (MSG, 5.97) (CNNLoc+, 3.55) (Ours, 2.99)
			};
			
			\addplot+[draw=none, fill=red!50] coordinates {
				(k-NN, 12.24) (CNNLoc, 12.37) (DLM, 16.38) (MSG, 14.16) (CNNLoc+, 9.69) (Ours, 7.56)
			};
			
			\legend{Median Error, 90th Percentile}
		\end{axis}
	\end{tikzpicture}
	\caption{Comparison of localization performance using median and 90th percentile errors. Our method achieves the lowest error in both metrics.}
	\label{fig:baseline-comparison}
\end{figure}


		%
		%
		%

\subsubsection{Comparison of Localization Methods}

Fig.~\ref{fig:baseline-comparison} illustrates the comparison of localization methods.
For large-scale deployments, traditional RSSI-based methods~\cite{torres2015comprehensive, jang2018indoor} deliver expected results, as these systems are optimized for such scenarios.
In contrast, CSI-based systems~\cite{arnold2019novel, liu2025graph}, which are typically designed for smaller-scale environments, fail to achieve comparable performance.
To address this limitation, we extended the approach proposed by~\cite{sandler2018mobilenetv2, jang2018indoor} and compared it with our system. 
Overall, our method demonstrates superior median and tail error performance, with a median error of 2.99 m and a tail error of 7.56 m. 

\subsubsection{Comparison of Pretraining Strategies}

\begin{figure}[b]
	\centering
	\begin{tikzpicture}
		\begin{axis}[
			width=0.95\linewidth,
			height=5.0cm,
			xlabel={Labeled Data Ratio (\%)},
			ylabel={Median Error},
			xtick={10,20,40,60,80,100},
			ymin=2.9, ymax=5.9,
			grid=both,
			legend style={at={(0.98,1.08)}, anchor=north east, font=\footnotesize},
			tick label style={font=\small},
			label style={font=\small},
			every axis plot/.append style={thick}
			]
			
			\addplot+[mark=*, dashed, color=gray] coordinates {
				(10, 5.69) (20, 5.04) (40, 4.34) (60, 3.94) (80, 3.59) (100, 3.40)
			};
			\addlegendentry{Random}
			
			\addplot+[mark=diamond, color=teal!70!black] coordinates {
				(10, 5.84) (20, 4.88) (40, 4.37) (60, 4.26) (80, 3.75) (100, 3.733)
			};
			\addlegendentry{Tri.}
			
			\addplot+[mark=square*, color=orange!80!black] coordinates {
				(10, 5.51) (20, 4.52) (40, 4.37) (60, 3.87) (80, 3.73) (100, 3.78)
			};
			\addlegendentry{GMAE}
			
			\addplot+[mark=triangle*, color=blue!60!black] coordinates {
				(10, 5.51) (20, 4.79) (40, 3.90) (60, 3.62) (80, 3.36) (100, 3.34)
			};
			\addlegendentry{GCL}
			
			\addplot+[mark=diamond*, color=teal!70!black] coordinates {
				(10, 5.14) (20, 4.24) (40, 3.95) (60, 3.66) (80, 3.54) (100, 3.43)
			};
			\addlegendentry{Pseudo.}
			
			\addplot+[mark=star, color=black] coordinates {
				(10, 4.73) (20, 4.15) (40, 3.71) (60, 3.25) (80, 3.28) (100, 3.18)
			};
			\addlegendentry{Ours}
			
			\node[font=\tiny, color=black] at (axis cs:10,4.58) {↑16.9\%};
			\node[font=\tiny, color=black] at (axis cs:20,4.00) {↑17.7\%};
			\node[font=\tiny, color=black] at (axis cs:40,3.56) {↑14.5\%};
			\node[font=\tiny, color=black] at (axis cs:60,3.10) {↑17.5\%};
			\node[font=\tiny, color=black] at (axis cs:80,3.13) {↑8.6\%};
			\node[font=\tiny, color=black] at (axis cs:100,3.03) {↑6.5\%};
			
		\end{axis}
	\end{tikzpicture}
	\caption{Comparative performance of pretraining methods under varying label ratios.}
	\label{fig:rep-learning-comparison}
\end{figure}
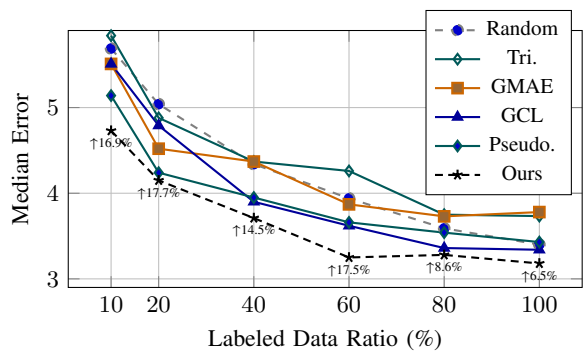

As shown in Fig.~\ref{fig:rep-learning-comparison}, we compare our method with the pre-training strategies mentioned above at various labeled dataset ratios. 
Empirical results demonstrate that pre-training strategies without specific adaptation to the localization task do not significantly improve performance and may even negatively impact it.
In contrast, our pre-training strategies consistently outperform the aforementioned approaches in enhancing localization task performance.


\subsubsection{Evaluation in Limited Data Scenarios}

\begin{figure}[h]
	\centering
	\includegraphics[width=1.0\columnwidth]{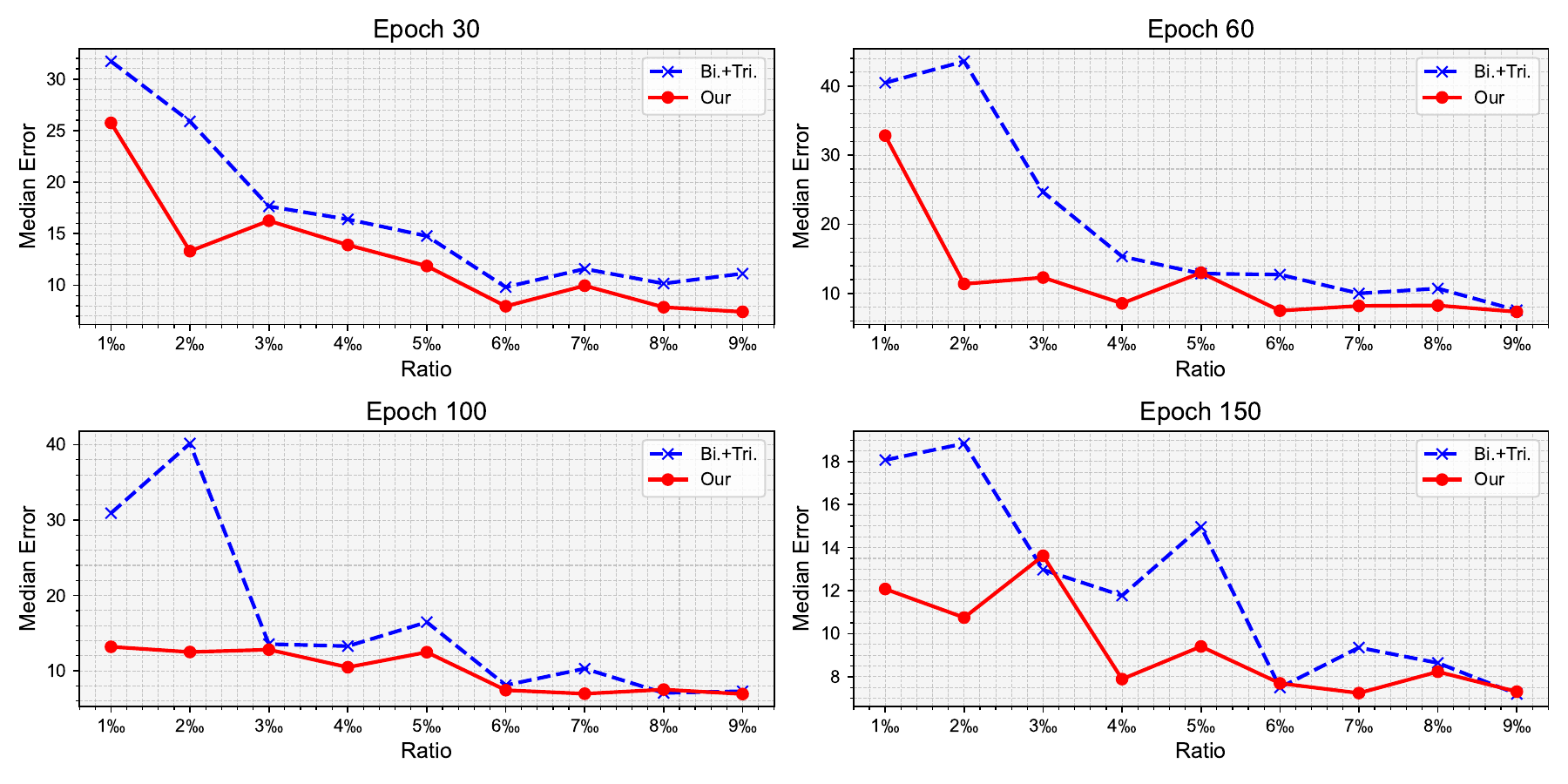}
	\caption{Performance comparison under limited labeled data with varying label ratios and pretraining epochs.}
\end{figure}

In this experiment, we evaluate and compare our method with Tri.+Bi~\cite{taner2023channel} under a setting with limited labeled data and various pretraining epoch configurations.
This evaluation is significant, as fewer labeled samples lead to lower site survey costs. 
Specifically, we evaluate both methods at epochs 30, 60, 100, and 150, using dataset ratios of 1-9 per thousand to highlight the advantages of our method. 
Our approach outperforms the Tri.+Bi. method, particularly in scenarios with limited labeled samples and fewer pretraining epochs.


\subsubsection{Building-level Leave-one-phone-out Evaluation}

\begin{figure}[h]
	\centering
	\includegraphics[width=1.0\columnwidth]{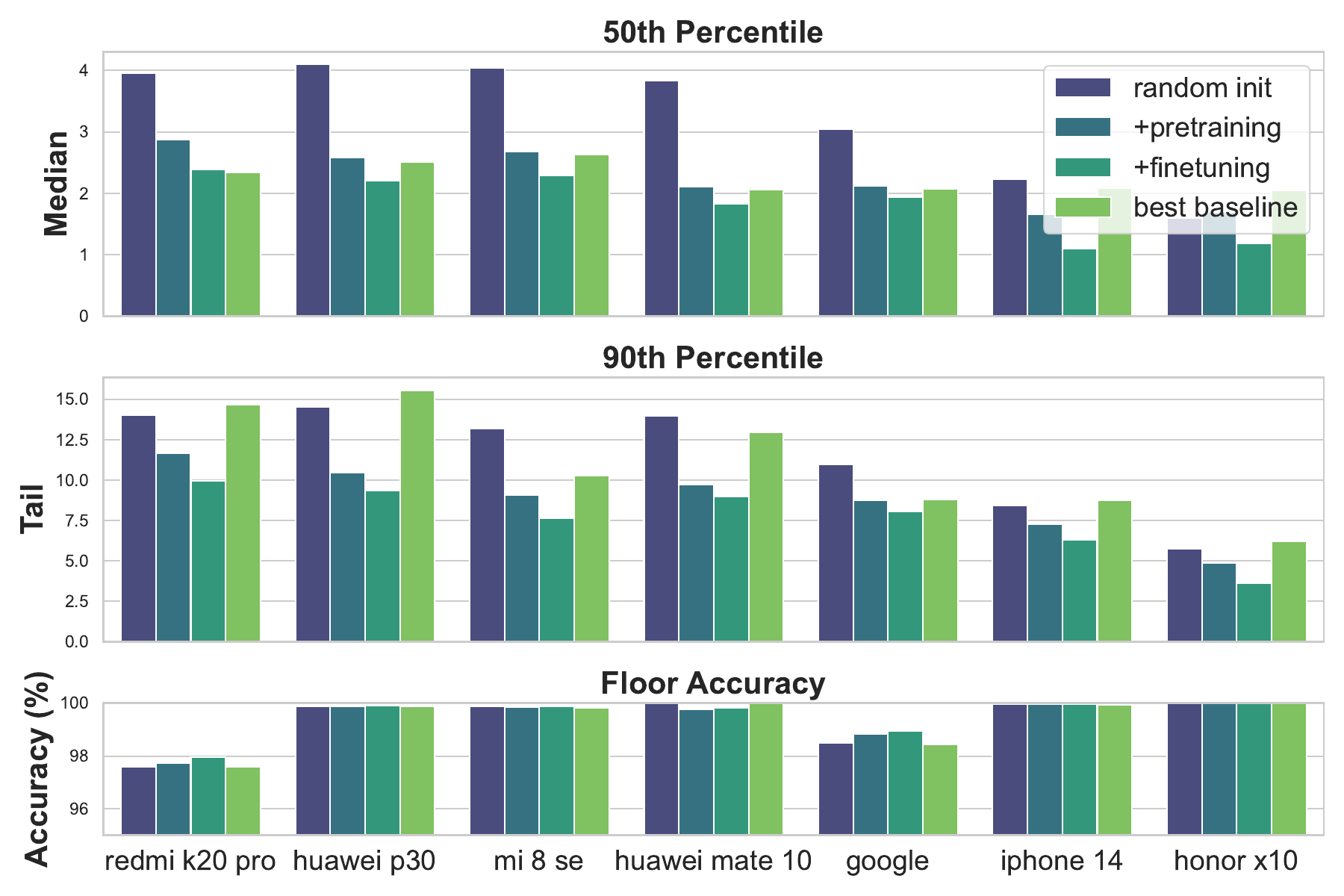}
	\caption{Performance comparison of the GLow model across different devices.}
	\label{fig:mobile}
\end{figure}

In this experiment, we evaluate the building-level leave-one-phone-out evaluation, where a specific type of mobile phone is excluded from the training dataset and used solely in the testing set. 

Our graph structure, which encodes heterogeneity and comprehensive CSI information, demonstrates limitations in cross-device generalizability, an issue we discussed earlier~\cite{vovk2005algorithmic}. 
However, as illustrated in Fig.~\ref{fig:mobile}, our pre-training and fine-tuning strategies effectively address this issue. 
A statistical analysis of errors across all tested mobile phones reveals that our method achieves a median score of 2.17m and a 90th percentile score of 8.93m, compared to the best baseline method, which scores a median of 2.28m and a 90th percentile of 12.16m. 
Moreover, GLow attains a floor accuracy of 99.49\%, which surpasses the 99.29\% achieved by the baseline method. 
Overall, GLow achieves an 18.7\% improvement in mean absolute error from 4.64m to 3.77m. 




\subsection{Detail Performance Experiment}
\label{detail}

\subsubsection{Ablation Study of Spatiotemporal Pretraining}

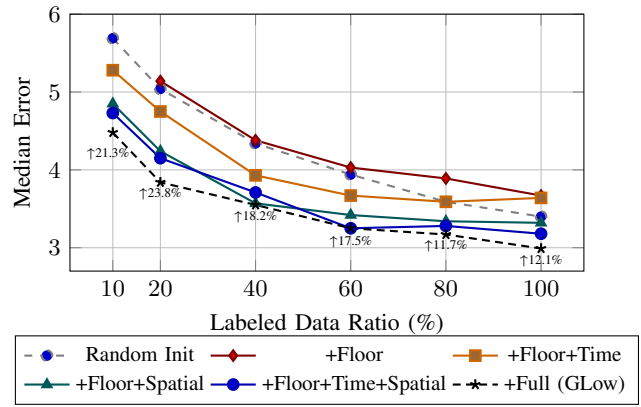
\begin{figure}[h]
	\centering
	\begin{tikzpicture}
		\begin{axis}[
			width=0.95\linewidth,
			height=5.0cm,
			xlabel={Labeled Data Ratio (\%)},
			ylabel={Median Error},
			xtick={10,20,40,60,80,100},
			ymin=2.7, ymax=6,
			grid=both,
			legend style={
				at={(0.5,-0.25)}, anchor=north, 
				legend columns=3, 
				font=\footnotesize
			},
			tick label style={font=\small},
			label style={font=\small},
			every axis plot/.append style={thick}
			]
			
			\addplot+[mark=*, dashed, color=gray] coordinates {
				(10, 5.69) (20, 5.04) (40, 4.34) (60, 3.94) (80, 3.59) (100, 3.40)
			};
			\addlegendentry{Random Init}
			
			\addplot+[mark=diamond*, color=black!50!red] coordinates {
				(20, 5.14) (40, 4.38) (60, 4.03) (80, 3.89) (100, 3.67)
			};
			\addlegendentry{+Floor}
			
			\addplot+[mark=square*, color=orange!80!black] coordinates {
				(10, 5.28) (20, 4.75) (40, 3.93) (60, 3.67) (80, 3.59) (100, 3.64)
			};
			\addlegendentry{+Floor+Time}
			
			\addplot+[mark=triangle*, color=teal!70!black] coordinates {
				(10, 4.85) (20, 4.24) (40, 3.57) (60, 3.42) (80, 3.34) (100, 3.32)
			};
			\addlegendentry{+Floor+Spatial}
			
			\addplot+[mark=*, color=blue!70!black] coordinates {
				(10, 4.73) (20, 4.15) (40, 3.71) (60, 3.25) (80, 3.28) (100, 3.18)
			};
			\addlegendentry{+Floor+Time+Spatial}
			
			\addplot+[mark=star, color=black] coordinates {
				(10, 4.48) (20, 3.84) (40, 3.55) (60, 3.25) (80, 3.17) (100, 2.99)
			};
			\addlegendentry{+Full (GLow)}
			
			\node[font=\tiny, color=black] at (axis cs:9,4.21) {↑21.3\%};
			\node[font=\tiny, color=black] at (axis cs:20,3.69) {↑23.8\%};
			\node[font=\tiny, color=black] at (axis cs:40,3.40) {↑18.2\%};
			\node[font=\tiny, color=black] at (axis cs:60,3.08) {↑17.5\%};
			\node[font=\tiny, color=black] at (axis cs:80,3.02) {↑11.7\%};
			\node[font=\tiny, color=black] at (axis cs:100,2.84) {↑12.1\%};
			
		\end{axis}
	\end{tikzpicture}
	\caption{Ablation study of Glow components across different labeled data ratios. }
	\label{fig:glow-ablation}
\end{figure}

As depicted in Fig.~\ref{fig:glow-ablation}, we analyze the impact of various prior information on fine-tuning performance through spatiotemporal pre-training. 
Initially, training with only pseudo floor labels shows no improvement. 
Next, we integrate temporal and spatial dimensions alongside pseudo floor labels. We observe that temporal constraints enhance performance when training samples account for less than 60\% of the total data, but the benefits plateau as the dataset size increases. 
In contrast, spatial constraints consistently improve performance across all sample sizes, although the marginal gains diminish with larger datasets. 
Finally, by incorporating confidence-aware fine-tuning, we achieve optimal performance.

\subsubsection{Sensitivity Analysis of Pretraining Parameters}

\begin{figure}[h]
	\centering
	\includegraphics[width=1.0\columnwidth]{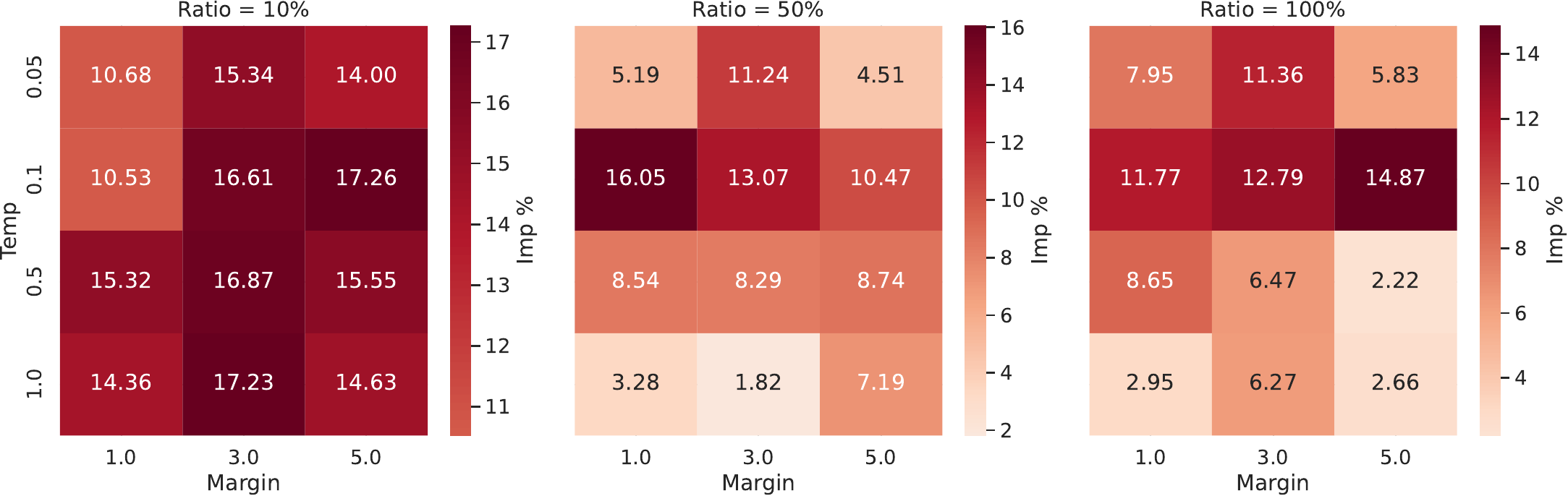}
	\caption{Heatmaps of performance improvement (\%) relative to random initialization under varying pretraining hyperparameters (temperature and margin) and labeled data ratios.}
	\label{fig:heatmap}
\end{figure}

In this experiment, we evaluate the impact of pre-training parameters, specifically margin and temperature, on fine-tuning performance. 
We systematically vary these parameters across different pre-training configurations and fine-tune the resulting models using varying ratios of the training dataset. 
As illustrated in Fig.~\ref{fig:heatmap}, we present heatmap visualizations of the improvement percentages (\textit{Imp\%}), which quantify the median error localization performance gains achieved through pre-training compared to models trained from scratch. 

When the training dataset ratio is low (e.g., 10\%), parameter selection does not play a critical role, although a margin of 3.0 demonstrates better performance.
As the training dataset ratio increases (e.g., 50\% or 100\%), the temperature parameter becomes the dominant factor influencing fine-tuning performance.
Among the tested configurations, a temperature of 0.1 consistently yields the best results.
This suggests that, with more fine-tuning data available, the fine-grained control provided by the temperature parameter becomes increasingly important, surpassing the impact of margin optimization. 

\subsubsection{Impact of Pre-training Epochs on Fine-tuning}

\begin{figure}[h]
	\centering
	\begin{tikzpicture}
		\begin{axis}[
			width=1.0\columnwidth,
			height=5cm,
			xlabel={Epoch},
			ylabel={Median Error},
			xtick={0,10,30,60,100,150},
			ymin=3.0, ymax=6.2,
			grid=both,
			legend style={
				at={(1.07,0.97)},
				anchor=north east,
				legend columns=3,
				font=\footnotesize,
				/tikz/every even column/.append style={column sep=4pt}
			},
			legend image post style={sharp plot, line width=1pt},
			tick label style={font=\small},
			label style={font=\small},
			every axis plot/.append style={thick},
			]
			
			\addplot [red, dashed, thick, opacity=0.4] coordinates {(0,5.69) (150,5.69)};
			\addplot [blue, dashed, thick, opacity=0.4] coordinates {(0,3.86) (150,3.86)};
			\addplot [green!60!black, dashed, thick, opacity=0.4] coordinates {(0,3.40) (150,3.40)};
			
			\addplot+[
			color=red,
			mark=*,
			nodes near coords,
			point meta=explicit symbolic,
			every node near coord/.append style={font=\tiny, color=red, yshift=4pt},
			] table[meta=label, x=epoch, y=error] {
				epoch error label
				0 5.69 {}
				10 5.86 {-2.98\%}
				30 4.93 {+13.35\%}
				60 4.84 {+14.93\%}
				100 4.63 {+18.62\%}
				150 4.73 {+16.87\%}
			};
			\addlegendentry{10\%}
			
			\addplot+[
			color=blue,
			mark=square*,
			nodes near coords,
			point meta=explicit symbolic,
			every node near coord/.append style={font=\tiny, color=blue, yshift=4pt},
			] table[meta=label, x=epoch, y=error] {
				epoch error label
				0 3.86 {}
				10 3.98 {-3.10\%}
				30 3.79 {+1.81\%}
				60 3.64 {+5.69\%}
				100 3.47 {+10.10\%}
				150 3.54 {+8.29\%}
			};
			\addlegendentry{50\%}
			
			\addplot+[
			color=green!60!black,
			mark=triangle*,
			nodes near coords,
			point meta=explicit symbolic,
			every node near coord/.append style={font=\tiny, color=green!60!black, yshift=4pt},
			] table[meta=label, x=epoch, y=error] {
				epoch error label
				0 3.40 {}
				10 3.51 {-3.23\%}
				30 3.35 {+1.47\%}
				60 3.30 {+2.94\%}
				100 3.15 {+7.35\%}
				150 3.18 {+6.47\%}
			};
			\addlegendentry{100\%}
			
		\end{axis}
	\end{tikzpicture}
	\caption{Effect of pretraining epochs on downstream median localization error under different labeled data ratios. Dashed lines indicate the no-pretraining baseline (epoch 0).}
	\label{fig:epoch-performance}
\end{figure}
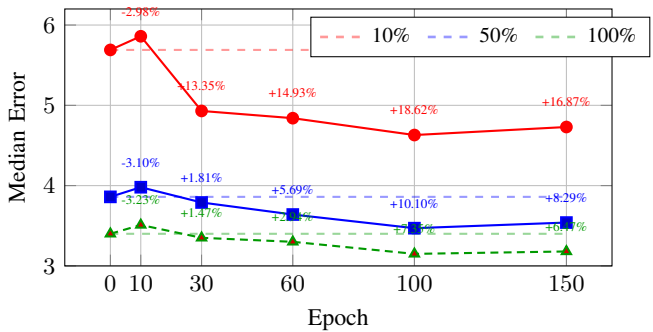

In this experiment, we examine how the number of pre-training epochs affects the fine-tuning performance of our model.
The model is pre-trained for varying durations (0, 10, 30, 60, 100, and 150 epochs) and subsequently fine-tuned on datasets comprising different proportions of the full training data (10\%, 50\%, and 100\%).
Fig.~\ref{fig:epoch-performance} presents the median error achieved under these conditions. 
Pre-training for 10 epochs results in higher median errors compared to training the model from scratch, suggesting that limited pre-training may impede the model's ability to generalize effectively during fine-tuning. 
As the number of pre-training epochs increases to 30, 60, and 100, there is a consistent decrease in median error across all data ratios, highlighting the positive impact of more extensive pre-training on fine-tuning performance. 
However, extending pre-training beyond 100 epochs to 150 yields no significant improvements, with only minor fluctuations, suggesting diminishing returns and a plateau in performance.

\subsubsection{Evaluation of Generalization Across Time}
To evaluate the generalization ability of our method over time, we consider two evaluation settings, each denoting a different temporal deployment scenario.
Specifically, the first investigates whether a single pretrained model can serve as a reusable backbone for fine-tuning on labeled data collected at later time points, reducing the need for repeated pretraining.
The second demonstrates the effectiveness of the proposed GLow design in addressing temporal generalization, by training the model on earlier data and directly evaluating it on future datasets without any further retraining. 

\begin{figure}[h]
	\centering
	\begin{tikzpicture}
		\begin{axis}[
			ybar,
			bar width=5pt,
			width=.9\linewidth,
			height=5cm,
			enlarge x limits=0.25,
			ylabel={Median Error},
			symbolic x coords={January, May, July},
			xtick=data,
			xtick style={draw=none},
			ymin=2, ymax=9,
			legend style={
				at={(0.03,1.07)},
				anchor=north west,
				font=\footnotesize,
				legend columns=3,
				/tikz/every even column/.append style={column sep=4pt}
			},
			ylabel style={font=\small},
			tick label style={font=\small},
			nodes near coords,
			nodes near coords align={vertical},
			every node near coord/.append style={font=\tiny, /pgf/number format/fixed},
			]
			
			\addplot+[draw=none, fill=red!30] coordinates {
				(January,5.69) (May,5.36) (July,6.83)
			};
			
			\addplot+[draw=none, fill=red!80] coordinates {
				(January,4.73) (May,4.28) (July,5.15)
			};
			
			\addplot+[draw=none, fill=blue!30] coordinates {
				(January,3.86) (May,3.57) (July,4.74)
			};
			
			\addplot+[draw=none, fill=blue!80] coordinates {
				(January,3.54) (May,3.06) (July,3.78)
			};
			
			\addplot+[draw=none, fill=gray!30] coordinates {
				(January,3.40) (May,2.71) (July,3.54)
			};
			
			\addplot+[draw=none, fill=black!50!] coordinates {
				(January,3.18) (May,2.53) (July,3.14)
			};
			
			\legend{
				10\% NP, 10\% +P, 
				50\% NP, 50\% +P, 
				100\% NP, 100\% +P
			}
		\end{axis}
	\end{tikzpicture}
	\caption{Median localization errors across three time periods under varying supervision ratios, with and without spatiotemporal pretraining. ``NP'' denotes training without pretraining; ``+P'' indicates spatiotemporal pretraining.}
	\label{fig:temporal-abs-error}
\end{figure}
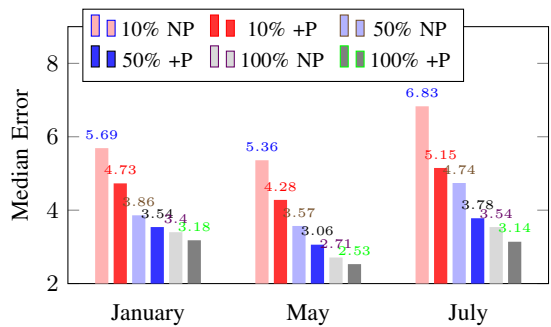

\textbf{Fine-Tuning on Future Data.} In this part, we evaluate the performance of the spatiotemporal pre-trained model over different time periods.
It is advantageous for the pre-trained model to be trained once and then utilized for fine-tuning at any subsequent time.
Therefore, we collected datasets in January (30k samples), May (19k samples), and July (40k samples), spanning half a year, to assess the temporal robustness of the pre-trained model. 
As shown in Fig.~\ref{fig:temporal-abs-error}, the pre-trained model consistently enhances the performance in downstream fine-tuning tasks, demonstrating robustness over time.

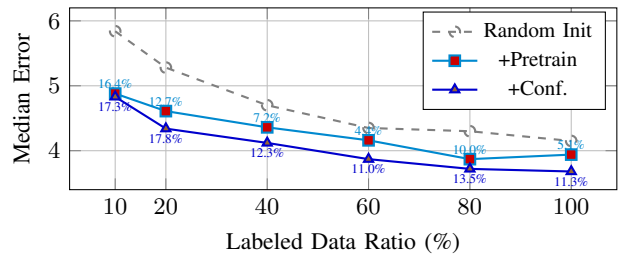
\begin{figure}[h]
	\centering
	\begin{tikzpicture}
		\begin{axis}[
			width=1.0\columnwidth,
			height=4cm,
			xlabel={Labeled Data Ratio (\%)},
			ylabel={Median Error},
			ymin=3.4, ymax=6.2,
			xtick={10,20,40,60,80,100},
			grid=both,
			tick label style={font=\small},
			label style={font=\small},
			legend style={at={(0.98,0.98)}, anchor=north east, font=\footnotesize},
			every axis plot/.append style={thick},
			]
			
			\addplot+[mark=o, dashed, color=gray] coordinates {
				(10, 5.84) (20, 5.28) (40, 4.70)
				(60, 4.35) (80, 4.30) (100, 4.15)
			};
			\addlegendentry{Random Init}
			
			\addplot+[mark=square*, color=cyan!70!blue] coordinates {
				(10, 4.88) (20, 4.61) (40, 4.36)
				(60, 4.16) (80, 3.87) (100, 3.94)
			};
			\addlegendentry{+Pretrain}
			
			\addplot+[mark=triangle*, color=blue!80!black] coordinates {
				(10, 4.83) (20, 4.34) (40, 4.12)
				(60, 3.87) (80, 3.72) (100, 3.68)
			};
			\addlegendentry{+Conf.}
			
			\node[font=\tiny, color=cyan!70!blue] at (axis cs:10,5.03) {16.4\%};
			\node[font=\tiny, color=cyan!70!blue] at (axis cs:20,4.76) {12.7\%};
			\node[font=\tiny, color=cyan!70!blue] at (axis cs:40,4.51) {7.2\%};
			\node[font=\tiny, color=cyan!70!blue] at (axis cs:60,4.31) {4.4\%};
			\node[font=\tiny, color=cyan!70!blue] at (axis cs:80,4.02) {10.0\%};
			\node[font=\tiny, color=cyan!70!blue] at (axis cs:100,4.09) {5.1\%};

			\node[font=\tiny, color=blue!80!black] at (axis cs:10,4.68) {17.3\%};
			\node[font=\tiny, color=blue!80!black] at (axis cs:20,4.19) {17.8\%};
			\node[font=\tiny, color=blue!80!black] at (axis cs:40,3.97) {12.3\%};
			\node[font=\tiny, color=blue!80!black] at (axis cs:60,3.72) {11.0\%};
			\node[font=\tiny, color=blue!80!black] at (axis cs:80,3.57) {13.5\%};
			\node[font=\tiny, color=blue!80!black] at (axis cs:100,3.53) {11.3\%};
			
		\end{axis}
	\end{tikzpicture}
	\caption{Median localization error and relative improvement (\%) four months post fine-tuning under varying labeled data ratios. Improvement rates are annotated.}
	\label{fig:longterm-eval}
\end{figure}

\textbf{Generalization to Future Data.} In this experiment, we evaluate the performance of the GLow component in the context of cross-temporal challenges in fingerprint-based localization. 
We specifically train the model using data collected in January and assess its generalization capability on data from May, approximately four months later.
We compare three training strategies: (i) training from scratch, (ii) pretraining followed by fine-tuning, and (iii) our full method with confidence-aware fine-tuning. 
The median localization error for each method, across various data ratios, is shown in Fig.~\ref{fig:longterm-eval}.
Overall, the results suggest that spatiotemporal pretraining and confidence-aware fine-tuning improve robustness across the temporal dimension.

\subsubsection{Confidence Estimation Evaluation}
\begin{table}[h]
	\centering
	\caption{Comparative Pearson Correlation Coefficients Across Different Labeled Training Set Ratios}
	\resizebox{0.9\columnwidth}{!}{
		\begin{tabularx}{\columnwidth}{@{}l *{6}{>{\centering\arraybackslash}X}@{}} 
			\toprule
			\textbf{Ratio} & 10\% & 20\% & 40\% & 60\% & 80\% & 100\% \\
			\midrule
			Pearson X & 0.45 & 0.49 & 0.52 & 0.54 & 0.56 & 0.59 \\
			Pearson Y & 0.51 & 0.54 & 0.63 & 0.66 & 0.71 & 0.70 \\
			\bottomrule
		\end{tabularx}
		\label{tab:pearson_correlation}
	}
\end{table}

We explore the benefits of our confidence estimation by comparing the Pearson Correlation Coefficient (PCC) between confidence score and error. 
The PCC, which ranges from \(-1\) to \(1\), quantifies the strength of a linear relationship, with values closer to \(1\) indicating a strong positive correlation. 
As shown in Tab.~\ref{tab:pearson_correlation}, the confidence estimation demonstrates a moderate to strong correlation with the error. 

\subsubsection{Time Efficiency Evaluation}

As indicated in Tab.~\ref{tab:method_comparison}, the graph-based method has lower storage costs and faster processing speed compared to the vector-based method. 
When used as input to the network, the vector-based encode inherently demands more computational resources and is less effective for parallel processing due to increased GPU memory usage, while the graph-based method offers faster processing speed. 
We evaluate these results based on averages from over 30,000 real-world samples. 
Tests on the NVIDIA GeForce RTX 4090 show that graph encoding processes 4k samples in just 0.11 seconds.

\section{Limitations and Future Work}
\label{section:limitations}
We conclude with a discussion of the limitations of our current design and ideas for future work:

\begin{itemize}
	\item Our work leverages landmark-based fingerprint collection methods, similar to those in~\cite{hu2022experience, hu2023wisdom}, to cover large-scale localization spaces. While effective, these methods still demand significant human effort. 
	We believe our approach can be integrated with solutions like SLAM-enabled robots~\cite{ayyalasomayajula2020deep} and domain-adversarial techniques~\cite{li2024train, chiu2025graph} for constructing and maintaining fingerprint databases, which will be explored in future work. 
	\item Similar to previous work~\cite{he2023sencom, liu2024wifo}, we unified these channel modes through data fitting. However, we believe that exploring the underlying principles behind these patterns offers a promising direction for more robust CSI feature extraction.
	\item Successful experiences~\cite{devlin2018bert, oquab2023dinov2} with pretraining on unlabeled datasets demonstrate that combining multiple auxiliary tasks is effective in unlocking the potential of unlabeled data. We believe that further integration of multiple auxiliary tasks in the future represents a promising direction. 
	\item Although our pretraining strategy has demonstrated improvements under limited labeled data, its performance still falls short of the precision required for real-world deployment.  We believe that achieving higher precision in localization under few-shot scenarios is an important and promising direction.  
\end{itemize}


\begin{table}[t]
	\centering
	\caption{Comparison of Graph-Based and Vector-Based Methods}
	\resizebox{0.9\columnwidth}{!}{
		\begin{tabular}{l c c c}
			\toprule
			\textbf{Method} & \textbf{Storage (units/GB)} & \textbf{CPU (s/4k)} & \textbf{GPU (s/4k)} \\
			\midrule
			Vector-Based & 0.6k & - & 2.23 \\
			Graph-Based & 15.3k & 1.14 & 0.11 \\
			\bottomrule
		\end{tabular}
	}
	\label{tab:method_comparison}
\end{table}

\section{CONCLUSION}
\label{section:conclusion}
In this paper, we introduced insights from deploying a learning-based CSI localization system on a large-scale ISAC platform.
We identified two key aspects regarding the leveraging of unlabeled datasets and the heterogeneity of CSI data, which are often overlooked by traditional CSI-based localization systems.
We employed a novel graph structure to capture the heterogeneous CSI data.
Furthermore, through the design of spatiotemporal pre-training and confidence-aware fine-tuning strategies, our system achieved optimal performance. 
Our comprehensive evaluations, which included over 70,000 data points across 25,600 \(m^2\) spanning five floors and involving seven smartphone types, demonstrated the practical potential of GLow as a prototype localization system. 
We believe that a large-scale deployment perspective will bring new insights and inspire fresh thinking in the community. 


\small
\bibliographystyle{IEEEtran}
\bibliography{sample-base}


\ifCLASSOPTIONcaptionsoff
\newpage
\fi

\end{document}